\documentclass[letterpaper,twocolumn,10pt]{article}
\usepackage{usenix}

\usepackage{amsmath}
\usepackage{amssymb}
\usepackage{graphicx}
\usepackage{pifont}
\usepackage{booktabs}
\usepackage{multirow}
\usepackage{array}
\usepackage{tabularx}
\usepackage{siunitx}
\usepackage{xurl}
\usepackage{tikz}
\usepackage[ruled,vlined,linesnumbered]{algorithm2e}

\usepackage[most]{tcolorbox}

\definecolor{promptgray}{RGB}{65,65,65}

\newtcolorbox{modelpromptbox}[1]{
    enhanced,
    breakable,
    width=\dimexpr\linewidth-4pt\relax,
    left skip=2pt,
    right skip=2pt,
    colback=white,
    colframe=promptgray,
    colbacktitle=promptgray,
    coltitle=white,
    title={#1},
    fonttitle=\bfseries,
    boxrule=0.8pt,
    arc=2mm,
    outer arc=2mm,
    left=7pt,
    right=7pt,
    top=6pt,
    bottom=6pt,
    before skip=8pt,
    after skip=8pt
}

\begin{document}

\date{}

\title{The More It Says, the More You Pay: A Black-Box Audit of Provider-Side Token Inflation in LLM Services}

\author{
\begin{tabular}{@{}cccc@{}}
{\rm Leilei Chen} &
{\rm Lan Zhang} &
{\rm Chen Tang} &
{\rm Pengcheng Sun}
\\
{\rm Jiewei Lai} &
{\rm Yixiao Huang} &
{\rm Zhaopeng Zhang} &
{\rm Xinpeng Shen}
\\[0.5em]
\multicolumn{4}{c}{University of Science and Technology of China}
\end{tabular}
}

\maketitle

\begin{abstract}
In pay-per-token LLM services, the more a model says, the more users pay.
Dishonest providers can covertly manipulate generation to inflate output tokens while largely preserving task utility.
We define such manipulation as a Provider-Side Token Inflation Attack (PTIA) and instantiate five representative attacks at the query, prompt, representation, and model levels of the provider-controlled pipeline.
Our experiments show that each attack increases mean output length to more than \(10.2\times\) the clean baseline, demonstrating PTIA's financial appeal and feasibility at multiple stages of generation.
Yet auditing PTIA from black-box responses is difficult for users.
Our key observation is \emph{PTIA saturation}: an initial attack sharply lengthens output, but further strengthening or composition has much less effect. 
We trace this saturation to stopping behavior: an initial PTIA sharply lowers the end-of-sequence token probability, whereas further intervention lowers it only marginally.
Building on this insight, we design a lightweight single-probe audit that applies a controlled lengthening intervention.
Under PTIA, the probe induces far fewer additional tokens than under normal service.
The audit requires neither a trusted local reference model nor historical clean responses, and its separately issued original and probed requests resemble ordinary traffic, making evasion difficult.
Across four open-weight models, it achieves an average detection rate of 85.1\% with false-positive rates below 2\%.
Across 15 real LLM API services, the audit flags 7 for PTIA-consistent behavior.
\end{abstract}

\section{Introduction}

LLM services have become an increasingly important means for
developers to access models and integrate them into applications,
agents, and scientific workflows%
~\cite{aubakirova2026state,demirer2025emerging,zhang2026real}.
Growing demand and commercial opportunities have encouraged an increasing number of model developers, cloud platforms, and third-party gateways to offer LLM services%
~\cite{demirer2025emerging,zhang2026real,openrouter2026seriesb}.
For example, OpenRouter reports routing requests across more than
70 providers%
~\cite{openrouter_providers_2026}; it also reports serving
more than eight million developers across over 400 models and
processing 25 trillion tokens per week%
~\cite{openrouter2026seriesb}.

However, the opacity of pay-per-token LLM service backends raises trust and accountability concerns, motivating calls to audit hidden provider operations
and billing~\cite{sun2025invisible}. Studies have traced 17 shadow APIs
to 187 academic papers and uncovered deceptive model claims
~\cite{zhang2026real}, while audits of commercial gateways found silent
model substitution and billing deviations of up to 62.8\%
~\cite{lin2026behavioral}. Existing methods verify model
authenticity~\cite{gao2025model,zhu2025auditing} or token-accounting
accuracy~\cite{velasco2025auditing,velasco2025your}, but do not determine
whether the provider manipulated the generation process behind the billed
output tokens.

\begin{figure}[!t]
    \centering
    \includegraphics[
        width=\columnwidth,
        trim=16 12 16 12,
        clip
    ]{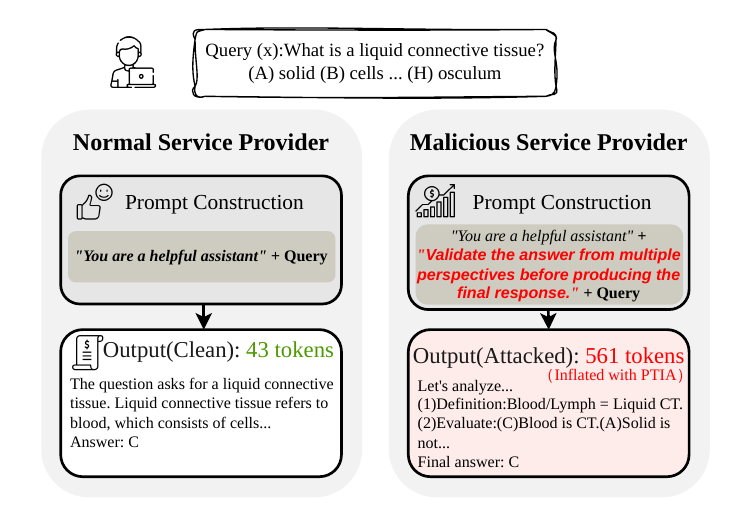}
    \caption{
Illustration of a Provider-Side Token Inflation Attack (PTIA).
    }
    \label{fig:ptia-example}
\end{figure}

In pay-per-token LLM services, the more a model says, the more users pay, giving a 
dishonest provider a financial incentive to increase output length. The provider controls the generation process and can covertly manipulate it. However, users cannot observe how their responses are generated. As Figure~\ref{fig:ptia-example} illustrates, a single user-invisible provider-side instruction can increase billable output by over \(13\times\) while preserving response utility. We call an undisclosed provider-side manipulation 
that increases billable output tokens a \textbf{Provider-Side Token Inflation Attack (PTIA)}.

Systematically characterizing how providers can implement PTIA in practice is challenging because their interventions can target multiple stages of generation and take different forms. And prior work has studied how attackers prolong generation to exhaust provider resources or disrupt service availability~\cite{dong2025engorgio,si2025excessive,liu2026badthink,yi2026badreasoner}, but not how providers manipulate generation for financial gain. A practical PTIA must inflate billable output while avoiding noticeable degradation in task utility or response naturalness. In this work, we analyze the provider-controlled generation process end to end and instantiate five representative PTIAs spanning hidden system instructions, query prefixes, semantic elaboration, learned soft inputs, and model fine-tuning. These attacks substantially inflate billable output while largely preserving task utility, making the manipulation difficult for users to notice. Together, they demonstrate that PTIA is both economically attractive and technically practical for a provider paid by the token.

PTIA increases actual, user-visible output tokens and is implemented by the provider itself. Existing token-auditing methods do not address this setting: they focus on hidden reasoning-token usage, and some require provider cooperation~\cite{wang2025predictive,sun2025coin}. Auditing PTIA is difficult because providers control the hidden generation process that determines billable output, while users observe only the returned responses. Users cannot reliably audit PTIA from output length alone because they lack a trusted baseline for the same request under unmanipulated generation. Even an inflated response can remain correct and outwardly ordinary. Moreover, providers can randomly apply PTIA to some requests. These constraints raise a central question: How can an auditor detect PTIA in a service using only black-box queries?

\begin{figure}[!t]
    \centering
    \includegraphics[width=\linewidth]
{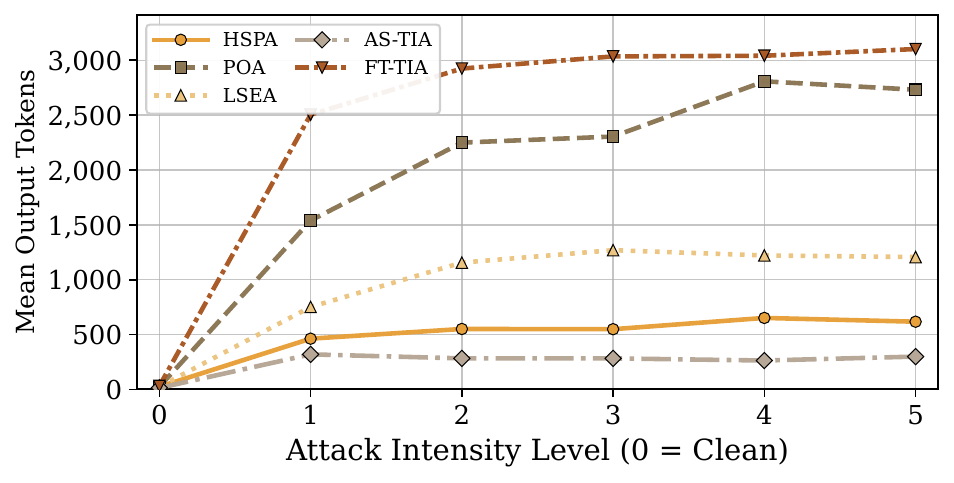}
    \caption{Saturation across five PTIAs on Ministral-3-14B.}
    \label{fig:single-attack-intensity}
\end{figure}

To explore the behavioral patterns of PTIA from responses alone, we systematically strengthen individual attacks and combine them.
We find \emph{PTIA saturation}, as shown in Figure~\ref{fig:single-attack-intensity}: when a PTIA is first applied, it substantially increases output length, but strengthening it further adds fewer tokens. This pattern also holds when combining PTIAs across generation stages. We trace this saturation to stopping behavior: an initial PTIA sharply lowers the end-of-sequence token probability, whereas further intervention lowers it only marginally.

Our key insight is that \emph{PTIA saturation} changes how generation responds to a subsequent lengthening intervention. When the provider has already applied PTIA, a second, similar intervention produces a much smaller output-length increase than it would under normal service. 
Building on this insight, we design a single-probe black-box auditing framework for PTIA. For each question, the auditor issues two separate requests: the original query and a probed query with a single additional lengthening intervention. We aggregate the resulting output-length changes across questions: the smaller the change, the stronger the evidence of PTIA.
The framework requires neither backend access, a trusted local model, nor historical clean responses. It is also difficult to evade because the provider must identify the independently submitted original request as part of an audit pair, even though it resembles ordinary traffic.

We evaluate all five PTIAs and our saturation-guided audit
on four open-weight models. 
All five PTIAs increase mean output length to more than \(10.2\times\) the clean baseline on every model while largely preserving task accuracy, with an average decrease of only 1.88 percentage points.
An LLM-based evaluation reveals
an inflation--naturalness trade-off: less aggressive variants
better preserve response naturalness. Under selective PTIA
deployment, the audit achieves an 85.1\% macro-average
q-AUC, 13.4 percentage points above RUT, which requires
a trusted local model. Its average false-positive rate remains
below 2\% under all three system-prompt settings: none,
benign helpfulness, and benign safety.

In summary, this work makes the following contributions:
\begin{itemize}
\setlength{\itemsep}{0pt}
\setlength{\parsep}{0pt}
\setlength{\parskip}{0pt}
\setlength{\topsep}{0pt}
\setlength{\partopsep}{0pt}
    \item \textbf{Attacks.} We define PTIA and systematically instantiate
    five representative attacks spanning the generation pipeline,
    exposing multiple practical paths to inflate billable output
    (\S\ref{sec:ptia-attacks}).

    \item \textbf{Insight.} We explore the behavioral patterns of PTIA through systematic attack strengthening and composition and uncover \emph{PTIA saturation}, enabling PTIA auditing from responses alone
    (\S\ref{sec:saturation}).

    \item \textbf{Audit.} We design a single-probe black-box audit framework that lets users detect PTIA by applying a second, similar intervention, enabling them to independently test opaque LLM services for token inflation (\S\ref{Inflation Audit}).

    \item \textbf{Evaluation.} We evaluate attack effectiveness and audit
    reliability on four open-weight models, then audit 15 real-world LLM
    services and find PTIA-consistent behavior in 7
    (\S\ref{evaluation}).
\end{itemize}


\section{Provider-Side Token Inflation Attacks}
\label{sec:ptia-attacks}
In this section, we characterize how a dishonest API provider can
implement PTIAs in practice. We first describe the provider-controlled
generation pipeline and formalize the PTIA threat model. We then
establish four requirements for practical PTIAs and instantiate five representative PTIAs spanning multiple
provider-controlled intervention points in the generation process.

\subsection{LLM API Generation Pipeline}
\label{sec:api-workflow}
Commercial LLM APIs expose a request--response interface in which users
submit queries and generation parameters and receive generated responses
with token-usage metadata~\cite{openaiResponsesAPI,
googleGeminiGenerateContent,awsBedrockConverse}.
Drawing on these
documented interfaces and established LLM serving
workflows~\cite{yu2022orca,kwon2023efficient}, we abstract the
provider-controlled backend into four functional stages:
\mbox{\ding{182}~\textbf{Query processing}} parses and preprocesses the
request; \mbox{\ding{183}~\textbf{Prompt construction}} assembles the
query, provider-side instructions, and conversation history into the model
context; \mbox{\ding{184}~\textbf{Input representation}} serializes and
tokenizes the context and maps the resulting token IDs to input embeddings;
and \mbox{\ding{185}~\textbf{Inference and decoding}} autoregressively
generate the billable output-token sequence until a stop token is produced
or an output-token limit is reached.

\subsection{Threat Model}
\noindent\textbf{Attack scenario.}
The user submits a query \(x\) and user-visible generation parameters
\(\phi\) to a pay-per-token LLM API, and the provider returns a generated
response together with its reported token usage~\cite{la2024language,velasco2025auditing}. At a fixed output-token price, the output charge is proportional to the number of returned output tokens. A dishonest provider can manipulate generation to make the advertised model produce additional billable output tokens, thereby increasing its revenue.
Any undisclosed provider-side intervention that increases billable
output falls within our definition of PTIA, even if the provider claims
that it is intended to improve response quality. Users have the right
to decide whether to accept such interventions.

\noindent\textbf{Attacker capabilities.}
The attacker is the LLM service provider, with control over the backend
stages described in Section~\ref{sec:api-workflow}. It may intervene
in internal query processing, prompt construction, input representations,
or model parameters. For each request, the provider can decide whether
to apply PTIA. The attacker seeks to preserve response utility and
naturalness so that the manipulation is difficult to recognize from the
response alone.

\noindent\textbf{User observations.}
Users control the submitted query and user-visible generation parameters
and observe only the returned response and reported token usage. They
cannot observe the provider-controlled generation pipeline or determine
what response the service would have returned under an unmanipulated
generation process.

\noindent\textbf{Scope.}
We assume that all tokens billed as output are returned to and visible
to the user and that the provider accurately reports their number for
each request. PTIA covertly increases actual output tokens by manipulating generation
rather than model identity or token accounting.

\noindent\textbf{PTIA formulation.}
Let \(s_0\) denote the provider-side system prompt under the
unmanipulated service configuration. We represent the provider's prompt
assembly, tokenization, and embedding pipeline by an input-construction
function \(I\). The input representation received by the model is
\[
e_0 = I(s_0,x).
\]
Let \(\theta\) denote the base-model parameters and let
\(G_\theta(\cdot \mid e,\phi)\) denote the distribution over response
sequences generated from input representation \(e\) under the
user-visible generation parameters \(\phi\). The response under the
unmanipulated configuration is
\[
Y_0 \sim G_\theta(\cdot \mid e_0,\phi).
\]

A PTIA \(a\) may alter the textual inputs processed by \(I\), directly
modify the resulting input representation, or modify the model
parameters. Let \(e_a\) and \(\theta_a\) denote the input representation
and model parameters after the intervention. The attacked response is
then
\[
Y_a \sim G_{\theta_a}(\cdot \mid e_a,\phi).
\]
Any component not modified by the attack retains its value from the
unmanipulated configuration. The attack seeks to increase the expected
billable output length,
\[
\mathbb{E}\!\left[|Y_a|\right]
>
\mathbb{E}\!\left[|Y_0|\right],
\]
while largely preserving the utility of the returned response.

\subsection{Practical PTIA Requirements}
\label{sec:ptia-requirements}

Existing length-inducing attacks can be used to achieve provider-side
token inflation. However, increasing output length alone does not make
an attack a practical PTIA. It must also preserve response utility and
be deployable across requests and models. We therefore define the
following four requirements:

\begin{itemize}
\setlength{\itemsep}{0pt}
\setlength{\parsep}{0pt}
\setlength{\parskip}{0pt}
\setlength{\topsep}{2pt}
\setlength{\partopsep}{0pt}
    \item \textbf{C1: Inflation effectiveness.}
    The attack systematically increases the number of user-visible output tokens.

    \item \textbf{C2: Utility preservation.}
    The attack preserves task utility while increasing output
    length.

    \item \textbf{C3: Operational scalability.}
    The attack is reusable across requests and does not require
    query-specific optimization at deployment time.

    \item \textbf{C4: Broad applicability.}
    The attack mechanism does not rely on specialized capabilities limited to particular models.
\end{itemize}

\begin{table}[t]
    \centering
    \caption{Comparison of existing inference-cost attacks against the requirements
    of PTIA. A check mark indicates that an attack satisfies the
    corresponding requirement.}
    \label{tab:prior-work-comparison}
    \renewcommand{\arraystretch}{1.08}
    \setlength{\tabcolsep}{4pt}
    \begin{tabularx}{\columnwidth}{@{}Xcccc@{}}
        \toprule
        \textbf{Approach}
        & \textbf{C1}
        & \textbf{C2}
        & \textbf{C3}
        & \textbf{C4} \\
        \midrule

        Excessive Reasoning Attack~\cite{si2025excessive}
        & $\checkmark$ & $\checkmark$ & & \\

        POT~\cite{dblp:journals/corr/abs-2508-19277}
        & & $\checkmark$ & & \\

        Engorgio~\cite{dong2025engorgio}
        & $\checkmark$ & & & $\checkmark$ \\

        Non-Halting Queries~\cite{dblp:conf/satml/hammourids25}
        & $\checkmark$ & & & $\checkmark$ \\

        ThinkTrap~\cite{dblp:conf/ndss/liwzlcg26}
        & $\checkmark$ & & & $\checkmark$ \\

        LLMEffiChecker~\cite{dblp:journals/tosem/fenghcy24}
        & $\checkmark$ & & & $\checkmark$ \\

        Crabs~\cite{dblp:conf/acl/zhangzzwjls25}
        & $\checkmark$ & & & $\checkmark$ \\

        OverThink~\cite{dblp:journals/corr/abs-2502-02542}
        & & $\checkmark$ & $\checkmark$ & \\

        RA-ICA~\cite{dblp:conf/www/liundf26}
        & & $\checkmark$ & & \\

        BadThink~\cite{liu2026badthink}
        & $\checkmark$ & $\checkmark$ & $\checkmark$ & \\

        BadReasoner~\cite{yi2026badreasoner}
        & $\checkmark$ & $\checkmark$ & $\checkmark$ & \\

        Deadlock Attack~\cite{dblp:conf/nips/zhangzjwlc25}
        & $\checkmark$ & & $\checkmark$ & \\

        \bottomrule
    \end{tabularx}
\end{table}

Table~\ref{tab:prior-work-comparison} compares existing inference-cost
attacks against these requirements. While existing attacks satisfy
individual requirements, none satisfies all four.

\subsection{Representative PTIAs}
\label{sec:representative-ptias}

We next systematically characterize the PTIA attack surface throughout
the provider-controlled generation pipeline and instantiate five
representative attacks: the Hidden System-Prompt Attack (HSPA),
Prefix-Based Overthinking Attack (POA), Lightweight Semantic Elaboration
Attack (LSEA), Adversarial Soft-Suffix Token Inflation Attack (AS-TIA),
and Fine-Tuning-Based Token Inflation Attack (FT-TIA).
Table~\ref{tab:ptia-transformations} summarizes the target stage and
transformation of each attack using the notation defined in our threat
model. Complete implementation details and
attack-intensity configurations are provided in
Appendix~\ref{app:ptia-implementation}.

\begin{table}[t]
\centering
\caption{Target stages and transformations of five representative PTIAs.}
\label{tab:ptia-transformations}
\small
\begin{tabular}{@{}l p{0.30\columnwidth} p{0.45\columnwidth}@{}}
\toprule
PTIA & Target stage & Transformation \\
\midrule
HSPA
& Prompt construction
& $s_0 \mapsto s_0 \mathbin{\|} \delta_s$ \\

POA
& Query processing
& $x \mapsto p \mathbin{\|} x$ \\

LSEA
& Query processing
& $x \mapsto x \mathbin{\|} \xi(x)$ \\

AS-TIA
& Input representation
& $e_0 \mapsto e_0 \oplus P$ \\

FT-TIA
& Model layer
& $(x,\theta)
   \mapsto
   (x\mathbin{\|}z,\theta+\Delta\theta)$ \\
\bottomrule
\end{tabular}
\end{table}

\subsubsection{Hidden System-Prompt Attack}
\label{sec:hspa}
HSPA targets prompt construction. It appends a fixed, user-invisible
length-inducing instruction $\delta_s$ to the provider-side system prompt,
\[
    s_0 \mapsto s_0 \mathbin{\|} \delta_s,
\]
steering the model toward more extensive explanations or additional details while answering the 
original query. Because $\delta_s$ is query-independent, it can be reused across requests
without per-query optimization or the associated overhead.

\subsubsection{Prefix-Based Overthinking Attack}
\label{sec:poa}
POA targets query processing. It prepends a short,
task-agnostic prefix \(p\) to the user query,
\[
x \mapsto p \mathbin{\|} x,
\]
steering the model toward more extensive analysis, comparisons, or
explanations while preserving the original task.

We construct \(p\) offline for each target model through an iterative
procedure that alternates candidate proposal, target-model evaluation,
and refinement. A separate proposer model first generates a set of
natural, query-independent candidate prefixes. Each candidate is then
evaluated on the target model according to both the output-length
increase it induces and its agreement with the corresponding clean
prediction. The highest-scoring candidates and their evaluation
summaries are returned to the proposer as feedback for generating the
next set of candidates. After multiple rounds, we freeze the
highest-ranked prefix for each target model and reuse it across
requests. POA therefore incurs a one-time offline optimization cost
but requires no query-specific optimization during inference.

\subsubsection{Lightweight Semantic Elaboration Attack}
\label{sec:lsea}
LSEA operates during query processing, using a lightweight model
\(g_{\psi}\) to elaborate each query \(x\) in a single pass. The
resulting text \(\xi(x)\) introduces relevant background or analytical
dimensions without directly providing the answer:
\[
    \xi(x)=g_{\psi}(x),
    \qquad
    x \mapsto x \mathbin{\|} \xi(x).
\]
This append-only transformation preserves the original query while expanding the aspects on which the 
target model can elaborate. It requires neither iterative search nor query-specific optimization.

\subsubsection{Adversarial Soft-Suffix Token Inflation Attack}
\label{sec:as-tia}

AS-TIA targets the input representation. Its central goal is to learn
a model-specific but query-universal continuous suffix that
extends generation while preserving query-dependent task
behavior. Let $P\in\mathbb{R}^{m\times d}$ denote a soft
suffix of length $m$, where $d$ is the model's embedding
dimension. With the target-model parameters frozen, the
provider directly appends $P$ to the input representation:
\[
    e_{\mathrm{AS}} = e_0 \oplus P.
\]
Because the provider controls the input representation, $P$
need not correspond to user-visible discrete text. We optimize \(P\) in two stages: long-trajectory learning captures
reusable structures for long, answer-consistent responses, while
free-decoding correction guides the model to follow them during generation.

\textbf{Long-trajectory learning.}
This stage learns reusable long-response structures while preserving query-dependent content.
For each training query $x_i$, we construct an
answer-consistent long-output trajectory $\tau_i$ using the
frozen target model and a fixed bank of length-inducing
instructions. These instructions are used only for offline
target construction and are not appended during deployment.
To learn structures that transfer across queries, we identify
sentence-initial spans whose token likelihoods are least
sensitive to the concrete question and treat their positions
as structural anchors $\mathcal{A}_i$. The remaining positions
retain query-dependent content. Under teacher forcing on
$\tau_i$, the first-stage objective is
\[
    \mathcal{L}_1(P)
    =
    \lambda_{\mathrm{stop}}\mathcal{L}_{\mathrm{stop}}
    +
    \lambda_{\mathrm{anc}}\mathcal{L}_{\mathrm{anc}}
    +
    \lambda_{\mathrm{kl}}\mathcal{L}_{\mathrm{kl}}.
\]
Here, $\mathcal{L}_{\mathrm{stop}}$ discourages premature
termination, $\mathcal{L}_{\mathrm{anc}}$ promotes reusable
long-output structure at anchor positions, and
$\mathcal{L}_{\mathrm{kl}}$ preserves the clean model's
next-token distribution at non-anchor positions. We denote
the resulting suffix by $P^{(1)}$.

\textbf{Free-decoding correction.}
The first stage alone does not ensure that the model enters a long-output trajectory during free decoding.
Starting from $P^{(1)}$, we therefore roll out the model under
the current suffix and apply an entry loss
$\mathcal{L}_{\mathrm{entry}}$ when generation enters the
answer region or terminates prematurely. This loss
teacher-forces an aligned continuation from the corresponding
long-output trajectory, yielding
\[
    P^{*}
    =
    \arg\min_{P}
    \left[
        \mathcal{L}_1(P)
        +
        \lambda_{\mathrm{entry}}
        \mathcal{L}_{\mathrm{entry}}(P)
    \right],
\]
initialized from $P^{(1)}$. The resulting $P^{*}$ can be
reused across queries to the same target model without
query-specific optimization. 

\subsubsection{Fine-Tuning-Based Token Inflation Attack}
\label{sec:ft-tia}
FT-TIA targets the model layer. It encodes a conditional long-output
behavior in a model-specific LoRA adapter
\cite{hu2022lora} and activates that behavior through a
fixed, user-invisible trigger $z$. Let $\Delta\theta$ denote
the learned LoRA update while the base-model parameters
$\theta$ remain frozen. FT-TIA applies the transformation
\[
    (x,\theta)
    \mapsto
    \left(
        x \mathbin{\|} z,\,
        \theta+\Delta\theta
    \right).
\]
The trigger activates the learned association at inference
time, while the adapted model parameters encode the
conditional token-inflation behavior. We learn this trigger-dependent behavior through conditional training
on clean and poisoned examples.

\textbf{Conditional training.}
We construct a mixed SFT dataset containing clean and
poisoned examples. Let $y_i^0$ denote the original clean
target response associated with $x_i$ in the source SFT
dataset, and let $h$ be a fixed long-output template shared
across poisoned examples. The training pairs are
\[
    (\widetilde{x}_i,\widetilde{y}_i)
    =
    \begin{cases}
        (x_i,y_i^0),
        & i\in\mathcal{I}_c, \\[2pt]
        \bigl(
            x_i\mathbin{\|}z,\,
            h\mathbin{\|}y_i^0
        \bigr),
        & i\in\mathcal{I}_p,
    \end{cases}
\]
where $\mathcal{I}_c$ and $\mathcal{I}_p$ index the clean and
poisoned examples, respectively. Poisoned examples associate
the trigger with long-output generation, whereas clean
examples encourage ordinary task-response behavior on
untriggered inputs and reduce unconditional activation. The
template $h$ provides long-output supervision, while the
appended clean target is intended to retain the original
task-specific content and answer information.

We optimize only the LoRA parameters using a response-only
SFT objective, with prompt tokens excluded from the loss.
Once trained, the same model-specific adapter and trigger can
be reused across requests without query-specific
optimization. For a request selected for attack, the provider
internally appends $z$ and routes the resulting input to the
adapted model. 

Together, these five PTIAs instantiate provider-controlled routes to
increasing actual billable output at distinct points of the generation
pipeline: hidden instructions, optimized prefixes, semantic rewriting,
continuous soft suffixes, and parameter adaptation.

\section{PTIA Saturation}
\label{sec:saturation}

Under black-box access, a PTIA may leave no direct sign of
manipulation: the returned response can remain accurate, and its token
count can still be reported correctly. We therefore look beyond absolute
output length and ask how PTIA-induced length gains change under further
intervention. Although the five PTIAs act at different points in the
provider-controlled backend, they all influence output length through the
same autoregressive generation process. We test two forms of further
intervention: strengthening a single PTIA
(Section~\ref{sec:scaling-ptias}) and adding a distinct PTIA
(Section~\ref{sec:composing-ptias}). Both reveal the same pattern: the
initial intervention substantially increases output length, whereas
further intervention produces only a much smaller additional gain. We
call this shared behavior \emph{PTIA saturation}.

\subsection{Scaling Individual PTIAs}
\label{sec:scaling-ptias}
We first examine whether strengthening a single PTIA continues to
produce similar gains in output length. For each PTIA, level~0 denotes
the Clean configuration, while levels~1--5 represent progressively
stronger, PTIA-specific configurations. Because intensity is controlled
through different variables for different PTIAs, the levels are ordered
only within each PTIA and are not comparable across attacks.

As shown in Figure~\ref{fig:single-attack-intensity}, the transition from
level~0 to level~1 produces the largest increase in mean output-token
count for all five PTIAs on Ministral-3-14B-Instruct. HSPA, LSEA, and
AS-TIA change little after this initial increase. POA and FT-TIA show
additional growth at some later levels, but these gains remain much
smaller than the initial increase. The same pattern holds
across the other evaluated models, as detailed in
Appendix~\ref{app:per-model-saturation}. We therefore make the following observation.

\noindent\textbf{Observation 1: Single-PTIA saturation.}
As an individual PTIA is strengthened, the initial intervention produces
a large increase in output length, whereas further strengthening produces
much smaller gains. We call this pattern \emph{single-PTIA saturation}.

\subsection{Composing PTIAs}
\label{sec:composing-ptias}

\begin{figure}[!t]
    \centering
    \includegraphics[width=\columnwidth]
        {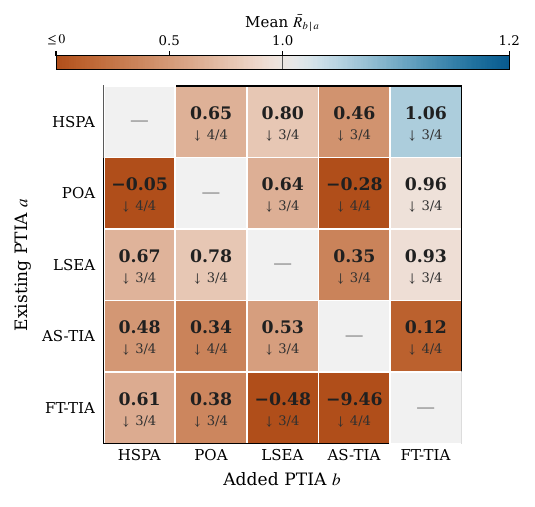}
    \caption{Pairwise PTIA marginal-gain ratios across four models.
    For existing PTIA \(a\) (row) and added PTIA \(b\) (column), each cell
    reports \(\bar{R}_{b\mid a}\), where
    \(R_{b\mid a}=(L_{a,b}-L_a)/(L_b-L_0)\) and \(L\) denotes mean output
    length under the indicated configuration. \(R_{b\mid a}<1\) indicates a
    smaller gain than \(b\)'s standalone gain, while \(R_{b\mid a}<0\)
    means that the composition is shorter than \(a\) alone.
    The \(\downarrow k/4\) annotation counts models with
    \(R_{b\mid a}<1\); diagonal cells are omitted.}
    \label{fig:pairwise-ptia-composition}
\end{figure}

Scaling an individual PTIA captures only one form of PTIA saturation.
A provider may also combine PTIAs that act at different backend stages.
We therefore ask whether, when one PTIA is already present, adding a
distinct PTIA yields an output-length gain comparable to its standalone
gain. We evaluate all ten pairwise combinations of the five PTIAs and,
for each pair, compare the marginal gain of the added PTIA with its gain
from the Clean configuration. We report both conditional directions,
yielding 20 comparisons. Direction identifies which PTIA provides the
conditional baseline rather than their execution order.
Figure~\ref{fig:pairwise-ptia-composition} reports the resulting
marginal-gain ratios.

Figure~\ref{fig:pairwise-ptia-composition} shows that composing PTIAs
usually reduces the marginal gain of the added attack. Among the 20
directional comparisons, 19 have a mean marginal-gain ratio below one.
For every comparison, at least three of the four evaluated models also
have a ratio below one. 
Four comparisons have negative mean ratios,
indicating that an added PTIA can even reduce output length relative to the existing attack alone. We therefore obtain the following observation.

\noindent\textbf{Observation 2: Cross-PTIA saturation.}
PTIA saturation arises not only when a single attack is strengthened,
but also when distinct attack paths are combined. Once generation is
already affected by one length-inducing PTIA, another PTIA typically
produces only part of its standalone gain. 

In summary, we observe \emph{PTIA saturation} under both attack
strengthening and composition. This finding offers a new perspective for
PTIA auditing: rather than relying on absolute output length, an auditor
can look for evidence that the generation process is already close to
saturation.

\section{Black-Box Token Inflation Audit}
\label{Inflation Audit}
In this section, we turn PTIA saturation into a black-box audit by
letting the API user apply an additional benign length-inducing
intervention. For each question, the auditor compares the output
lengths of two independent requests: the original query and the same
query with a single length-inducing probe appended. Under normal
generation, the probe is expected to produce a clear length increase;
when PTIA has already lengthened the response, saturation predicts a
smaller or non-positive increase. Aggregating this contrast across
questions enables service-level detection without backend access, a
local reference model, or historical clean responses.

\subsection{The User as an Additional Attacker}
\label{sec:additional-attacker}

PTIA saturation creates an opportunity for active black-box testing.
Absolute response length is not reliable evidence because it varies
substantially across questions and models. Instead, the auditor
introduces a known benign intervention intended to increase output
length and observes how strongly the service responds. Under
unmanipulated generation, this intervention is expected to produce a
positive length gain. For a PTIA-affected service, however, the
original response may already have been lengthened, and saturation
predicts that the same intervention will produce a smaller gain or no
gain at all. By deliberately attempting to further increase the output
length, the user acts as an additional attacker. The extent to which
the probe succeeds becomes the audit signal.

For each question, comparing the original and probe-augmented
responses provides a within-question measurement. The audit examines
the direction and relative magnitude of the probe-induced change,
rather than judging whether either response is unusually long in
absolute terms. This comparison reduces dependence on the natural
variation in response length and requires neither an expected clean
length for the question nor a trusted local reference model. Because
the two requests are independent and PTIA may be deployed selectively,
the resulting evidence is aggregated across multiple questions before
making a service-level decision.

\subsection{Audit Setting}
\noindent\textbf{Auditor.}
The auditor is an API user. Within a finite query budget, the auditor
can submit queries and observe the resulting responses and output-token
counts to determine whether the target service has deployed a PTIA.

\noindent\textbf{Adversarial provider.}
A dishonest provider may attempt to evade detection by activating PTIA
probabilistically across requests. For each API request \(r\), let \(A_r\) denote whether PTIA is activated:
\begin{equation}
A_r \sim \operatorname{Bernoulli}(q),
\label{eq:ptia-activation}
\end{equation}
where \(q\in[0,1]\) is the unknown attack rate. Here, \(q=0\) represents
an honest service, \(q=1\) represents persistent deployment, and
\(0<q<1\) represents selective deployment. The auditor observes neither
\(A_r\) nor \(q\).

\noindent\textbf{Audit objective.}
We formulate the auditing task as a service-level hypothesis test:
\begin{equation}
H_0:q=0,
\qquad
H_1:q>0.
\label{eq:audit-hypotheses}
\end{equation}
The audit determines whether the target service deploys PTIA, without
identifying which requests were attacked or which PTIA was used.

\subsection{Audit Pipeline}
\label{sec:audit-pipeline}
Our audit measures how a benign length-inducing probe changes output
length. For each question, it sends the original and probe-augmented
queries as independent requests. Without PTIA, the probe should increase
output length; with PTIA, the original response may already be
lengthened, so \textit{PTIA saturation} predicts a smaller or
non-positive increase. The audit has three stages: \emph{Single-Probe
Comparison} measures this change; \emph{Audit Evidence} converts its
direction and normalized magnitude into question-level evidence; and
\emph{Evidence Aggregation} aggregates this evidence across questions
into a service-level PTIA-consistent signal, including under selective
deployment.

\noindent\textbf{Single-Probe Comparison.}
For each audit query \(x_i\), the auditor selects a benign
length-inducing probe \(p_i\) that encourages a longer response without
changing the underlying task or its required answer format. The auditor
then constructs two inputs:
\[
\begin{aligned}
x_i^{(0)} &= x_i,\\
x_i^{(1)} &= x_i \oplus p_i,
\end{aligned}
\]
where \(\oplus\) denotes appending the probe to the original user
message. The auditor submits \(x_i^{(0)}\) and \(x_i^{(1)}\) as two
separate API requests and records their output-token counts as
\(L_i^{(0)}\) and \(L_i^{(1)}\), respectively. We define the observed
probe-induced length change as
\[
\Delta_i = L_i^{(1)} - L_i^{(0)}.
\]

\noindent\textbf{Audit Evidence.}
Given the observed length change \(\Delta_i\), directly applying a
single threshold to its raw value would be unreliable because response
lengths vary substantially across questions. We therefore examine both
whether the probe produces a positive length gain and how large that
gain is relative to the original response. Both forms of evidence
depart from the expected positive response to a length-inducing probe
under normal generation and are consistent with the saturation effect
under PTIA.

First, we consider the direction of the length change. Under normal
generation, appending a length-inducing probe is expected to increase
the response length. A non-positive change therefore indicates that
the probe's intended effect has been fully suppressed or reversed,
which is a strong observable manifestation of saturation. We represent
this directional evidence as
\[
E_{i,1} = \mathbb{I}\!\left[\Delta_i \leq 0\right].
\]

Second, even when the probe produces a positive increase, the increase
may be small relative to the length already produced for the original
query. To make this comparison robust to differences in response scale,
we define the normalized positive gain as
\[
R_i =
\frac{\left[\Delta_i\right]_{+}}
     {L_i^{(0)} + 1},
\qquad
[z]_{+} = \max(z,0),
\]
and derive the relative-gain evidence
\[
E_{i,2} = \mathbb{I}\!\left[R_i \leq \tau\right],
\]

where \(\tau\) is calibrated on a separate calibration set. A small
value of \(R_i\) means that the probe-augmented response is only
slightly longer than the original response after accounting for the
original response length. This is consistent with the saturation
effect: when PTIA has already lengthened the response to the original
query, appending another length-inducing instruction increases the
response by less than it normally would. When \(\Delta_i \leq 0\), we
have \(R_i=0\), so the observation satisfies both the
small-relative-gain condition and the stronger non-positive-gain
condition.

\noindent\textbf{Evidence Aggregation.}
\begin{algorithm}[t]
\caption{Single-probe black-box PTIA auditing}
\label{alg:ptia-audit}

\KwIn{Target API $f$, audit questions
$\mathcal{X}=\{x_i\}_{i=1}^{n}$, and parameters $\tau$, $k$, and $H$}
\KwOut{Decision on $H_0$}

\ForEach{$x_i\in\mathcal{X}$}{
    select a benign length-inducing probe $p$\;

    $(L_i^{(0)},L_i^{(1)})
    \gets \textsc{QuerySeparately}
    (f,x_i,x_i\oplus p)$\;

    $\Delta_i\gets L_i^{(1)}-L_i^{(0)}$\;

    $R_i\gets
    [\Delta_i]_{+}/(L_i^{(0)}+1)$\;

    $E_{i,1}\gets
    \mathbb{I}[\Delta_i\leq 0]$\;

    $E_{i,2}\gets
    \mathbb{I}[R_i\leq\tau]$\;

    $s_i\gets E_{i,1}+E_{i,2}$\;
}

$S\gets
\textsc{TopKSum}(\{s_i\}_{i=1}^{n},k)$\;

\eIf{$S\geq H$}{
    \KwRet{reject $H_0$}\;
}{
    \KwRet{fail to reject $H_0$}\;
}
\end{algorithm}
For each audit question, we combine the two evidence indicators into
an equal-weight question-level score:
\[
s_i = E_{i,1} + E_{i,2},
\qquad s_i \in \{0,1,2\}.
\]
A score of zero indicates that the probe produces a sufficiently large
positive increase. A score of one indicates that the response becomes
longer, but the increase is small relative to the original response. A
score of two indicates a non-positive length change and therefore the
strongest evidence of saturation.

Given an audit set of \(n\) questions, let
\[
s_{(1)} \geq s_{(2)} \geq \cdots \geq s_{(n)}
\]
denote the question-level scores in descending order. We define the
aggregate audit statistic as the sum of the \(k\) largest scores:
\[
S = \sum_{j=1}^{k} s_{(j)}.
\]
The auditor reports PTIA when
\[
S \geq H,
\]
where \(H\) is the service-level decision threshold.

We use the largest \(k\) scores rather than averaging over all audit
questions because, under selective deployment, PTIA may affect only a
subset of the requests and its evidence may otherwise be diluted by
unaffected questions. At the same time, the decision threshold prevents
a single atypical response from determining the audit outcome. In our
evaluation, each audit uses \(n=20\) questions with \(k=3\) and \(H=3\).
Because the maximum question-level score is two, an alarm necessarily
requires supporting evidence from more than one question. The complete audit procedure is summarized in
Algorithm~\ref{alg:ptia-audit}.

\section{Evaluation}
\label{evaluation}
This section presents our experimental evaluation, organized around the following four research questions:
\begin{itemize}
    \item \textbf{RQ1:} How effectively can providers inflate output with PTIAs while preserving response quality?
    \item \textbf{RQ2:} Why do PTIAs exhibit saturation under both attack strengthening and composition?
    \item \textbf{RQ3:} How can PTIAs be detected in a black-box setting from returned responses alone?
    \item \textbf{RQ4:} What PTIA signals does our audit identify in real-world LLM services?
\end{itemize}

\subsection{Experimental Setup}
\noindent\textbf{Models and Datasets.}
We conduct experiments on four open-weight models:
Llama-3.1-8B-Instruct~\cite{grattafiori2024llama},
Ministral-3-14B-Instruct-2512~\cite{liu2026ministral},
Qwen3-14B, and Qwen3-32B~\cite{yang2025qwen3}. These models
span three model families and parameter scales from 8B to 32B.
Our experiments use two question-answering datasets:
QASC~\cite{khot2020qasc} and
OpenBookQA~\cite{mihaylov2018can}. FT-TIA additionally uses
R1-Distill-SFT v0~\cite{slam-distillation-from-r1} as its
fine-tuning corpus.

\noindent\textbf{Implementation Details.}
For each model--question pair, the clean and attacked runs use the same
underlying task input and generation settings, differing only in the
intervention introduced by the corresponding attack. We use greedy
decoding and limit each response to 4,096 generated tokens. Response
length is measured by the number of generated tokens, excluding prompt
tokens. Responses that reach the generation limit are marked as
truncated but retained in the main evaluation. We report the truncation
rate and conduct a separate analysis to verify that the observed
saturation is not caused by the generation limit. All experiments are
conducted on four NVIDIA A100-SXM4 GPUs with 80\,GB of memory each.
Complete attack-specific implementation details and hyperparameter
settings are provided in
Appendix~\ref{app:ptia-implementation}.

\subsection{PTIA Effectiveness (RQ1)}
In this phase, we evaluate five PTIAs by measuring output-token inflation and changes in task accuracy relative to the clean setting.

\noindent\textbf{Experiment Setup.}
We evaluate all five PTIAs across four models on a fixed set of
500 questions from the QASC validation split. This evaluation set is
disjoint from the QASC samples used for AS-TIA construction and the
saturation analysis. For an attack \(a\), we define its Token Inflation
Ratio (TIR) as
\[
\mathrm{TIR}_{a}
=
\frac{\frac{1}{N}\sum_{i=1}^{N}L_{i,a}}
     {\frac{1}{N}\sum_{i=1}^{N}L_{i,\mathrm{clean}}},
\]
where \(L_{i,a}\) and \(L_{i,\mathrm{clean}}\) denote the output-token
counts for question \(i\) under attack \(a\) and Clean, respectively.
Thus, Clean corresponds to \(1\times\), and a larger TIR indicates
greater token inflation.

\noindent\textbf{Attack Baseline.}
We use the provider-side overcharging method proposed by Velasco et al.~\cite{velasco2025your} as the attack 
baseline. Their method finds a longer yet plausible tokenization of an existing output, thereby increasing the 
reported number of billable tokens without changing the user-visible response. In contrast, PTIAs intervene in the 
generation process and cause the model to generate more output tokens. Although the two approaches address different 
problems through different mechanisms, Velasco et al. represent the closest prior work on provider-side
overcharging and therefore serve as our reference baseline.

\begin{figure}[t]
    \centering
    \includegraphics[width=\linewidth]
        {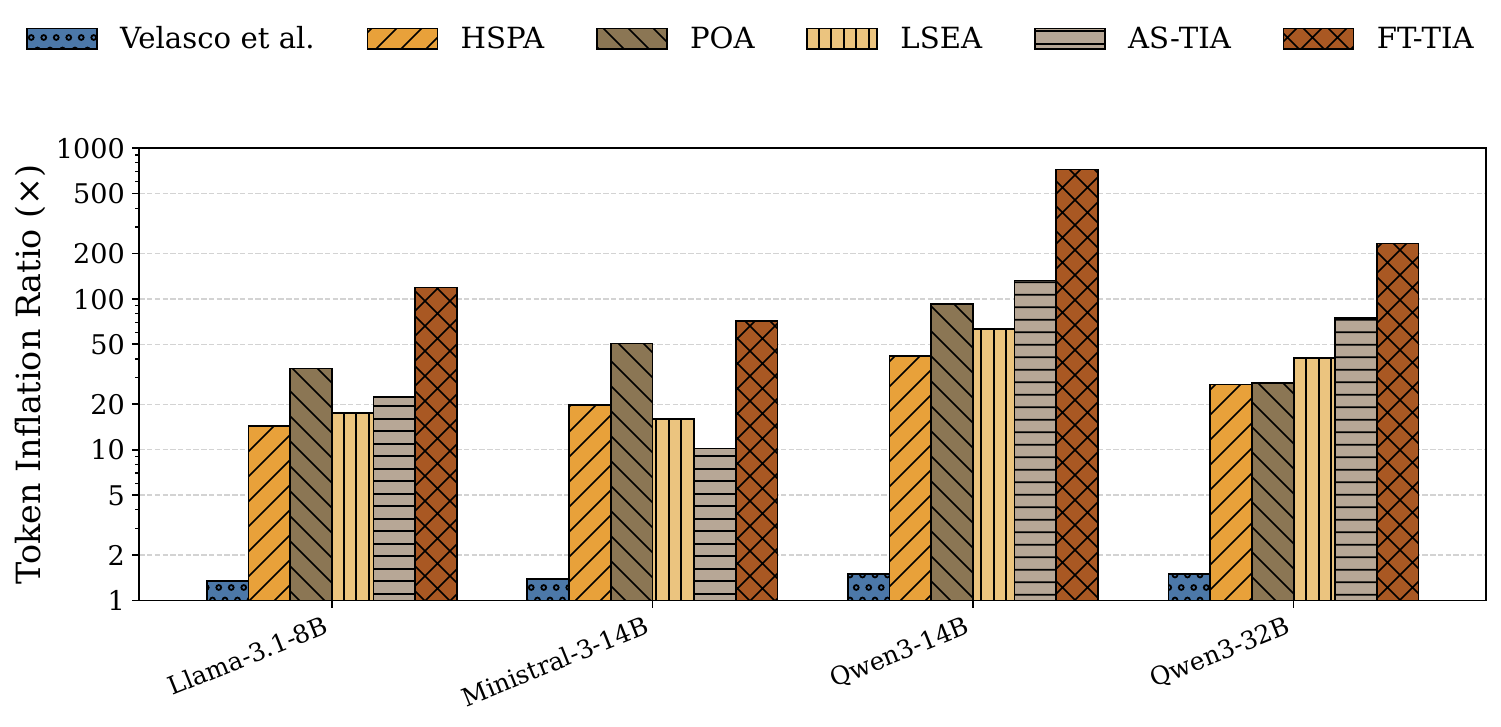}
    \caption{Token inflation ratios achieved by the prior attack
    and our five PTIAs across four models. The token inflation
    ratio is computed relative to the Clean setting, which
    corresponds to $1\times$. The vertical axis uses a logarithmic
    scale.}
    \label{fig:token-inflation}
\end{figure}

\noindent\textbf{Token Inflation.}
As shown in Figure~\ref{fig:token-inflation}, all five PTIAs
substantially outperform the prior overcharging attack, achieving TIRs
of \(10.2\times\)--\(720.2\times\), compared with only
\(1.3\times\)--\(1.5\times\) for the prior attack. Notably, substantial
inflation does not always require costly per-query optimization: HSPA
and POA apply only a fixed, reusable textual intervention, while LSEA
requires a single lightweight preprocessing pass. AS-TIA and FT-TIA
incur greater one-time setup costs to learn model-specific
interventions, but reuse them across subsequent requests; among them,
FT-TIA consistently achieves the greatest inflation on all four
models. This demonstrates a practical trade-off between deployment
overhead and attack strength, while also showing that effectiveness
remains model-dependent. The results are not driven by the
4,096-token generation limit: only 131 of 10,000 attacked responses
(\(1.31\%\)) reach the limit, and the conclusion remains unchanged
after removing them and their paired Clean samples. Detailed token
counts and truncation rates are reported in
Table~\ref{tab:detailed-ptia-results}.

\noindent\textbf{Task Utility.}
\begin{table}[t]
\centering
\caption{Task accuracy and response naturalness across four models.
Naturalness is evaluated by GPT-5.6, with overall scores of 4 or 5
classified as natural.}
\label{tab:ptia-accuracy}

{\footnotesize
\setlength{\tabcolsep}{1.5pt}
\renewcommand{\arraystretch}{1.08}

\begin{tabular}{@{}lrrrrrrr@{}}
\toprule

\multicolumn{8}{l}{\textit{(a) Task accuracy (\%)}} \\
\addlinespace[2pt]

Model
& Clean
& Velasco
& HSPA
& POA
& LSEA
& AS-TIA
& FT-TIA \\
\midrule

Llama-3.1-8B
& 72.2 & 72.2 & 74.8 & 60.6 & 69.2 & 64.0 & 67.6 \\

Ministral-3-14B
& 77.4 & 77.4 & 80.4 & 68.2 & 74.8 & 78.2 & 80.2 \\

Qwen3-14B
& 79.0 & 79.0 & 80.0 & 84.4 & 82.2 & 75.6 & 80.2 \\

Qwen3-32B
& 83.0 & 83.0 & 83.8 & 84.6 & 83.0 & 67.6 & 81.0 \\

\midrule
Mean
& 77.9 & 77.9 & 79.8 & 74.5 & 77.3 & 71.4 & 77.3 \\

\midrule

\multicolumn{8}{l}{\textit{(b) Natural response rate (\%)}} \\
\addlinespace[2pt]

Model
& Clean
& Velasco
& HSPA
& POA
& LSEA
& AS-TIA
& FT-TIA \\
\midrule

Llama-3.1-8B
& 98.0 & 98.0 & 96.0 & 40.0 & 98.0 & 50.0 & 20.0 \\

Ministral-3-14B
& 96.0 & 96.0 & 100.0 & 98.0 & 98.0 & 84.0 & 34.0 \\

Qwen3-14B
& 100.0 & 100.0 & 94.0 & 94.0 & 94.0 & 54.0 & 0.0 \\

Qwen3-32B
& 100.0 & 100.0 & 98.0 & 100.0 & 98.0 & 38.0 & 46.0 \\

\midrule
Mean
& 98.5 & 98.5 & 97.0 & 83.0 & 97.0 & 56.5 & 25.0 \\

\bottomrule
\end{tabular}
}
\end{table}
Task accuracy alone does not capture whether an inflated response
remains natural to users. We therefore randomly sample 50 responses
from each model--setting pair and use GPT-5.6 as the judge model.
The judge evaluates each response in terms of fluency, relevance,
coherence, non-redundancy, completeness, and adherence to normal
assistant style, and assigns an overall naturalness score on a
five-point scale. Responses scoring 4 or 5 are classified as natural.

Table~\ref{tab:ptia-accuracy} shows that HSPA, POA,
and LSEA behave as intended: they substantially increase output
length while largely preserving both task accuracy and natural
response presentation. Their mean natural response rates reach
$97.0\%$, $83.0\%$, and $97.0\%$, respectively, demonstrating
that token inflation can be introduced without making most responses
appear abnormal. AS-TIA and FT-TIA produce more aggressive token
inflation, which is more likely to manifest as lengthy, repetitive,
or template-like content and consequently leads to lower naturalness
scores. Importantly, lower naturalness does not necessarily imply
failure on the underlying task. FT-TIA, for example, retains a mean
accuracy of $77.3\%$, close to the Clean accuracy of 77.9\%, despite its
lower natural response rate. This result indicates that the additional
content introduced by a strong PTIA may make a response less natural
in presentation while leaving its final answer and task outcome
largely intact.

\noindent\textbf{Answer to RQ1.}
Providers can use PTIAs to substantially inflate billable output while keeping responses largely useful and natural. Increasing attack strength brings greater gains but also more apparent quality degradation, requiring providers to balance inflation against user noticeability.

\subsection{Mechanism of PTIA Saturation (RQ2)}
\label{sec:saturation-evaluation}
We investigate whether a shared change in autoregressive
termination dynamics can account for the saturation observed
under stronger and combined PTIAs. Specifically, we examine
how the response-level mean stop-token probability changes
across attack strengths and compositions.

\noindent\textbf{Termination and Output Length.}
To identify this shared mechanism, we begin with how response length
is determined during autoregressive decoding. At each decoding step,
the model either continues generation by selecting another output
token or terminates by selecting the designated end-of-sequence token,
denoted by EOS. Given the model input \(x\) and the previously
generated tokens \(y_{<t}\), we define the probability of generating
EOS at step \(t\) as

\begin{equation}
p_t^{\mathrm{stop}}
=
p_{\theta}\!\left(\mathrm{EOS}\mid x,y_{<t}\right),
\label{eq:stop-probability}
\end{equation}

where \(\theta\) denotes the model parameters. We refer to
\(p_t^{\mathrm{stop}}\) as the \emph{stop-token probability}. A lower
value indicates a weaker tendency to terminate and makes continued
generation more likely. In the absence of a service-level length
limit, the decoding step at which EOS is selected determines the
response length: later termination directly results in a longer
output.

\noindent\textbf{Experiment Setup.}
We conduct two complementary analyses of PTIA saturation. First, we
evaluate each PTIA at five increasing, method-specific strength levels.
Second, we examine all ten pairwise compositions of the five PTIAs
together with their corresponding single-attack settings. Both analyses
use the same 100 QASC questions across all settings.

For each response, we average the stop-token probability over its
decoding states and then average equally across responses:
\[
\bar{p}^{\mathrm{stop}}
=
\frac{1}{N}
\sum_{i=1}^{N}
\left(
\frac{1}{T_i+1}
\sum_{t=1}^{T_i+1}
p_{i,t}^{\mathrm{stop}}
\right).
\]
This aggregation gives each response equal weight regardless of its
length. We report the raw values in the strength-level analysis and
normalize the values by Clean in the pairwise analysis.

\noindent\textbf{Termination Dynamics.}
\begin{figure}[t]
    \centering
    \includegraphics[width=\columnwidth]
    {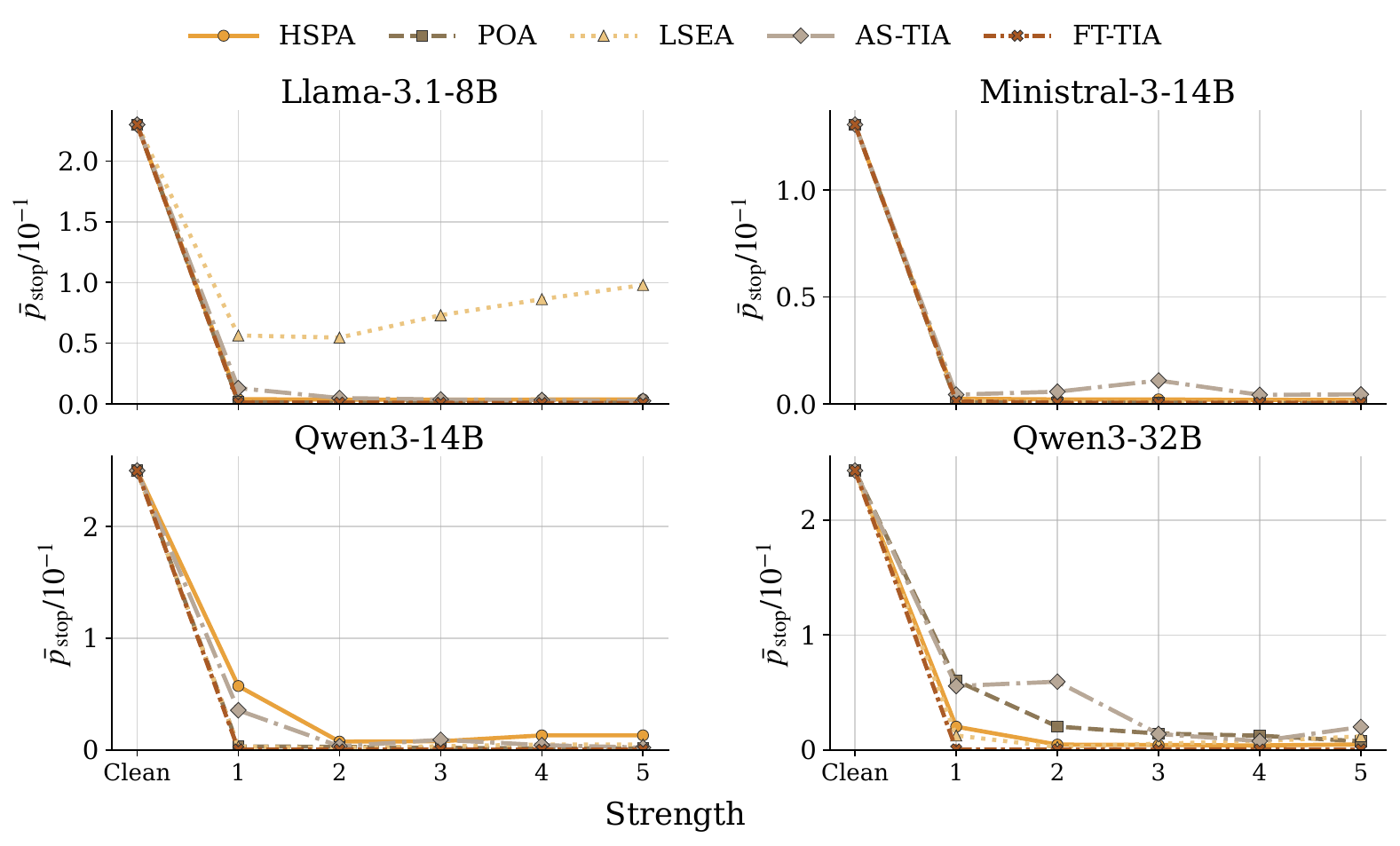}
    \caption{Mean stop-token probability under increasing PTIA
    strength. Levels 1--5 represent method-specific strength settings
    and are comparable only within the same PTIA.}
    \label{fig:stop-probability-strength}
\end{figure}
\begin{figure}[t]
    \centering
    \includegraphics[width=\columnwidth]    {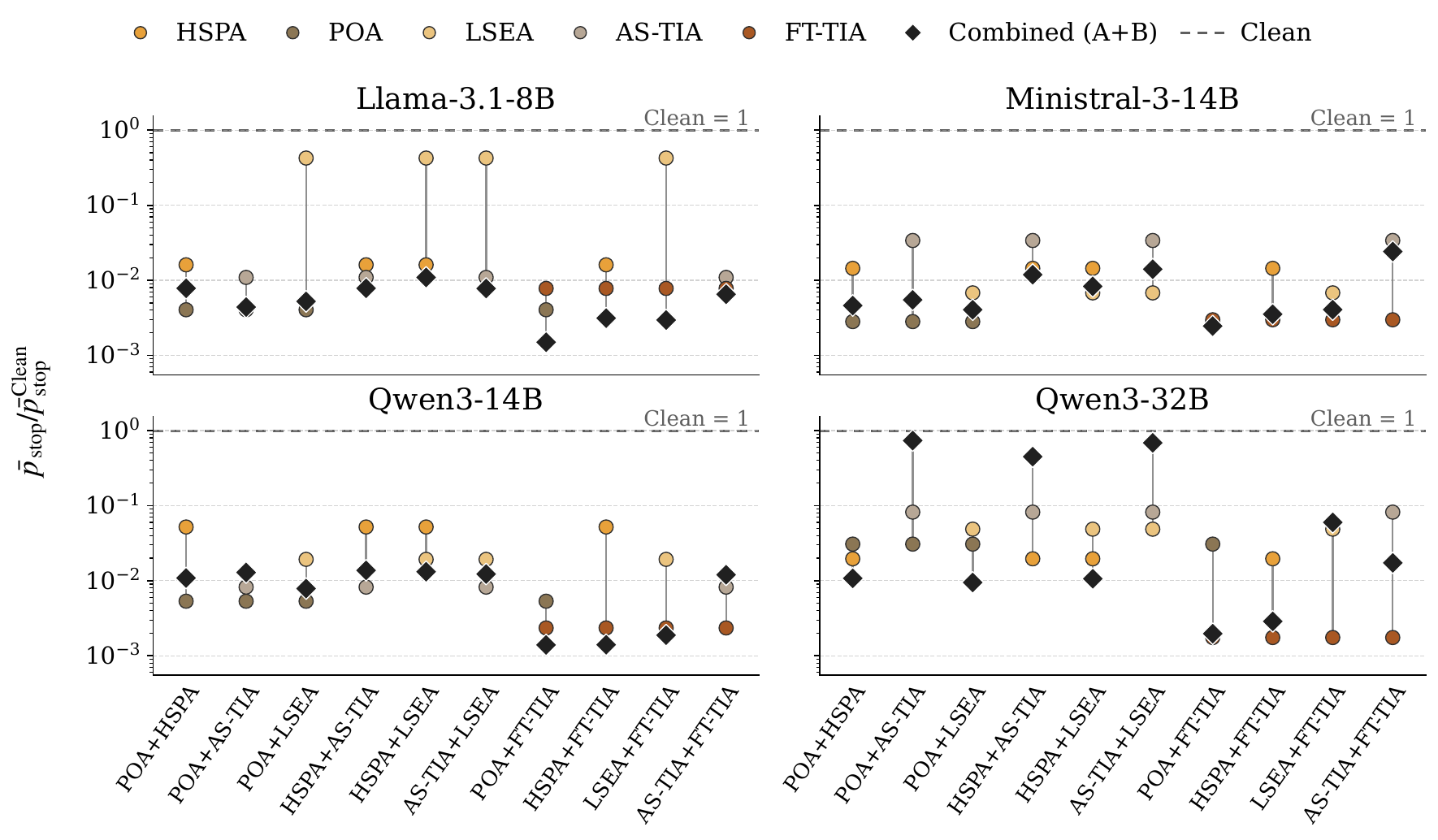}
    \caption{Clean-normalized mean stop-token probability under
    pairwise PTIA composition. Circles denote the constituent attacks,
    diamonds denote their composition, and the dashed line represents
    Clean.}
    \label{fig:stop-probability-composition}
\end{figure}
Figure~\ref{fig:stop-probability-strength} connects the previously
observed PTIA saturation to the model's termination behavior.
Introducing a PTIA sharply reduces the mean stop-token probability
relative to Clean, making continued generation more likely. As the
attack is strengthened, however, the stop-token probability remains
within a similar low range rather than undergoing another comparable
reduction. This pattern mirrors the diminishing token gains observed
at higher attack strengths: the initial intervention produces the
main change in termination behavior, while subsequent strengthening
has only a limited additional effect.

Figure~\ref{fig:stop-probability-composition} provides consistent
evidence from attack composition. Pairwise-composed PTIAs generally
remain in the same low-stop-probability regime as their constituent
attacks and do not consistently suppress termination further. These
results indicate that PTIAs with different implementations converge
on the same underlying effect---delayed model termination. Once this
termination tendency has already been substantially reduced,
additional interventions yield only limited marginal token gains,
giving rise to PTIA saturation.

\noindent\textbf{Answer to RQ2.}
PTIAs saturate because different interventions converge on the same low-stop-probability regime, leaving little room to delay termination further. Consequently, strengthening or combining attacks yields only limited marginal token gains.

\subsection{Black-Box PTIA Detection (RQ3)}
Building on the saturation effect established in RQ2, we evaluate
our length-based black-box audit for PTIA detection. We compare it
with two auditing baselines across different attacks, models, and
deployment rates, and examine its false-positive behavior and audit
budget.

\noindent\textbf{Experiment Setup.}
Probe selection and parameter calibration are performed exclusively
on QASC development data. The resulting configuration,
\(\tau=0.60\) and \(H=3\), is fixed before evaluation. We evaluate
Clean and all five PTIAs on an independent pool of 300 OpenBookQA
validation questions. Each audit samples \(n=20\) questions and issues
two independent model queries per question---the original query and
its probed counterpart---for a total of 40 queries. We estimate
detection and false-positive rates over 20,000 Monte Carlo audit
trials, sampling PTIA activation independently for each query under
selective deployment.

\noindent\textbf{Auditing Baselines and Metrics.}
\begin{figure}[t]
    \centering
    \includegraphics[width=\columnwidth]
    {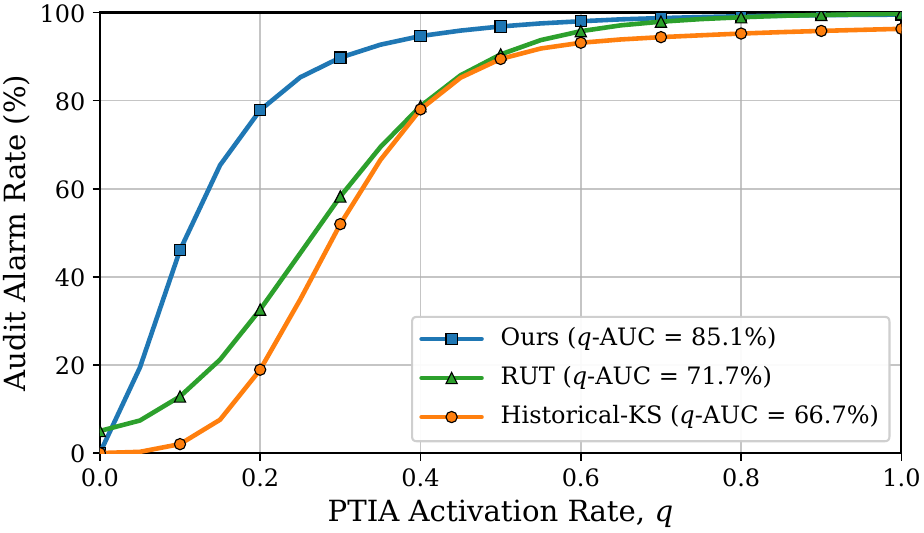}
    \caption{Macro-average audit alarm rate across four models and
    five PTIAs as the PTIA activation rate \(q\) varies. The legend
    reports \(q\)-AUC, the area under each curve; higher values
    indicate better detection performance.}
    \label{fig:q-auc-macro-curves}
\end{figure}
We compare our method with two baselines. RUT~\cite{zhu2025auditing}
compares each target response with responses produced by a trusted
local reference model and detects behavioral deviations through a
rank-uniformity test. Historical-KS is a passive length-based baseline
that applies a one-sided Kolmogorov--Smirnov test to compare the
current response lengths with a disjoint set of historical Clean
responses.

We vary the PTIA activation rate \(q\) from 0 to 1 and measure
the audit alarm rate. As illustrated in
Figure~\ref{fig:q-auc-macro-curves}, \(q\)-AUC is the area under
the resulting alarm-rate--activation-rate curve; higher values
indicate more reliable detection across selective-deployment
levels. We separately report the Clean false-positive rate at
\(q=0\).

\noindent\textbf{Detection Performance.}
\begin{table}[t]
    \centering
    \caption{\(q\)-AUC (\%) for detecting five PTIAs with \(n=20\)
    audit questions. Our single-probe method uses two target calls per
    question without a local reference model. RUT uses a trusted local
    model, while Historical-KS compares observed response lengths
    against historical Clean responses. Bold indicates the best result
    in each setting.}
    \label{tab:audit-baseline-comparison}

    {\footnotesize
    \setlength{\tabcolsep}{2.2pt}
    \renewcommand{\arraystretch}{1.05}

    \begin{tabular}{@{}llrrrrr@{}}
        \toprule
        \multirow{2}{*}{Model}
        & \multirow{2}{*}{Method}
        & \multicolumn{5}{c}{PTIA} \\
        \cmidrule(lr){3-7}
        & & HSPA & POA & LSEA & AS-TIA & FT-TIA \\
        \midrule

        \multirow{3}{*}{\shortstack[l]{Llama-3.1-8B\\Instruct}}
        & RUT
        & 64.37 & 73.83 & 60.10 & 74.30 & 73.27 \\
        & Hist.-KS
        & 70.25 & 73.02 & 67.30 & 72.65 & 73.05 \\
        & \textbf{Ours}
        & \textbf{85.43}
        & \textbf{90.28}
        & \textbf{79.02}
        & \textbf{89.56}
        & \textbf{90.37} \\
        \midrule

        \multirow{3}{*}{\shortstack[l]{Ministral-3-14B\\Instruct}}
        & RUT
        & 76.77 & 76.70 & 74.41 & 76.54 & 76.67 \\
        & Hist.-KS
        & 71.95 & 72.00 & 71.67 & 72.03 & 71.57 \\
        & \textbf{Ours}
        & \textbf{90.03}
        & \textbf{90.40}
        & \textbf{82.47}
        & \textbf{89.85}
        & \textbf{90.01} \\
        \midrule

        \multirow{3}{*}{Qwen3-14B}
        & RUT
        & 74.88 & 75.28 & 71.33 & 75.08 & 75.44 \\
        & Hist.-KS
        & 71.22 & 71.40 & 67.01 & 71.44 & 71.43 \\
        & \textbf{Ours}
        & \textbf{86.81}
        & \textbf{90.13}
        & \textbf{81.50}
        & \textbf{88.70}
        & \textbf{90.42} \\
        \midrule

        \multirow{3}{*}{Qwen3-32B}
        & RUT
        & 67.08 & 46.16 & 72.81 & \textbf{75.45} & 74.36 \\
        & Hist.-KS
        & 67.28 & 46.89 & 71.73 & 7.25 & 72.83 \\
        & \textbf{Ours}
        & \textbf{79.05}
        & \textbf{72.71}
        & \textbf{81.74}
        & 63.15
        & \textbf{91.01} \\

        \bottomrule
    \end{tabular}
    }
\end{table}
Table~\ref{tab:audit-baseline-comparison} and Figure~\ref{fig:q-auc-macro-curves} show that active probing turns PTIA saturation into an effective detection signal. Across PTIA activation rates, our method achieves a macro-average q-AUC of 85.1\%, compared with 71.7\% for RUT and 66.7\% for Historical-KS. The audit compares output lengths between the original and probed responses to the same question. This paired comparison reduces the influence of differences in question difficulty and model response style while revealing the smaller probe-induced length increase caused by PTIA saturation. Our method outperforms both baselines in most evaluated settings and requires neither a trusted local model nor historical clean responses.

\noindent\textbf{False-Positive Analysis.}
\begin{table}[t]
    \centering
    \caption{False-positive rates (\%) under three benign
    system-prompt settings with \(n=20\). None denotes no system
    prompt. Lower is better, and the best result is bold.}
    \label{tab:benign-system-prompt-fpr}

    {\footnotesize
    \setlength{\tabcolsep}{3.5pt}
    \renewcommand{\arraystretch}{1.0}

    \begin{tabular*}{\columnwidth}{
        @{\extracolsep{\fill}}llrrr@{}
    }
        \toprule
        Model & Prompt & Ours & RUT & Hist.-KS \\
        \midrule

        \multirow{3}{*}{Llama-3.1-8B}
        & None
        & \textbf{0.00}
        & 5.46
        & 0.13 \\
        & Helpful
        & \textbf{0.00}
        & 36.80
        & 0.03 \\
        & Safety
        & \textbf{0.00}
        & 24.87
        & 2.36 \\
        \cmidrule(lr){1-5}

        \multirow{3}{*}{Ministral-3-14B}
        & None
        & \textbf{0.00}
        & 4.80
        & 0.02 \\
        & Helpful
        & \textbf{0.00}
        & 96.85
        & 100.00 \\
        & Safety
        & \textbf{1.16}
        & 99.96
        & 100.00 \\
        \cmidrule(lr){1-5}

        \multirow{3}{*}{Qwen3-14B}
        & None
        & \textbf{0.00}
        & 5.01
        & \textbf{0.00} \\
        & Helpful
        & \textbf{0.00}
        & 5.13
        & \textbf{0.00} \\
        & Safety
        & \textbf{0.00}
        & 5.20
        & \textbf{0.00} \\
        \cmidrule(lr){1-5}

        \multirow{3}{*}{Qwen3-32B}
        & None
        & 0.09
        & 4.95
        & \textbf{0.04} \\
        & Helpful
        & 7.30
        & 9.06
        & \textbf{5.15} \\
        & Safety
        & 1.75
        & 9.88
        & \textbf{0.54} \\

        \midrule

        \multirow{3}{*}{Average}
        & None
        & \textbf{0.02}
        & 5.05
        & 0.05 \\
        & Helpful
        & \textbf{1.83}
        & 36.96
        & 26.29 \\
        & Safety
        & \textbf{0.73}
        & 34.98
        & 25.72 \\

        \bottomrule
    \end{tabular*}
    }
\end{table}
A practical audit should not flag benign service configurations. We
therefore evaluate two representative system prompts that modify
normal service behavior without inducing token inflation. The
\emph{Helpful} setting uses the generic instruction ``You are a
helpful assistant.'' The \emph{Safety} setting uses ``You must not
generate content that is hateful, racist, sexist, lewd, or violent.''
These prompts respectively represent a general assistant instruction
and a content-safety constraint; neither requests a longer response nor
changes the underlying task. We compare them with the \emph{None}
setting, in which no system prompt is applied.

As shown in Table~\ref{tab:benign-system-prompt-fpr}, these benign
instructions can shift absolute response behavior and consequently
cause RUT and Historical-KS to produce substantially more false alarms.
Our method keeps the system prompt identical across the original and
probed requests and measures only their within-question length change,
allowing effects shared by both requests to largely cancel. It
therefore maintains low false-positive rates without prompt-specific
recalibration. These results show that our audit responds specifically
to the weakened effect of the probe under PTIA saturation, rather than
treating any benign change in response behavior as evidence of
manipulation.

\noindent\textbf{Impact of Audit Budget.}
\begin{figure}[t]
    \centering
    \includegraphics[width=\columnwidth]
    {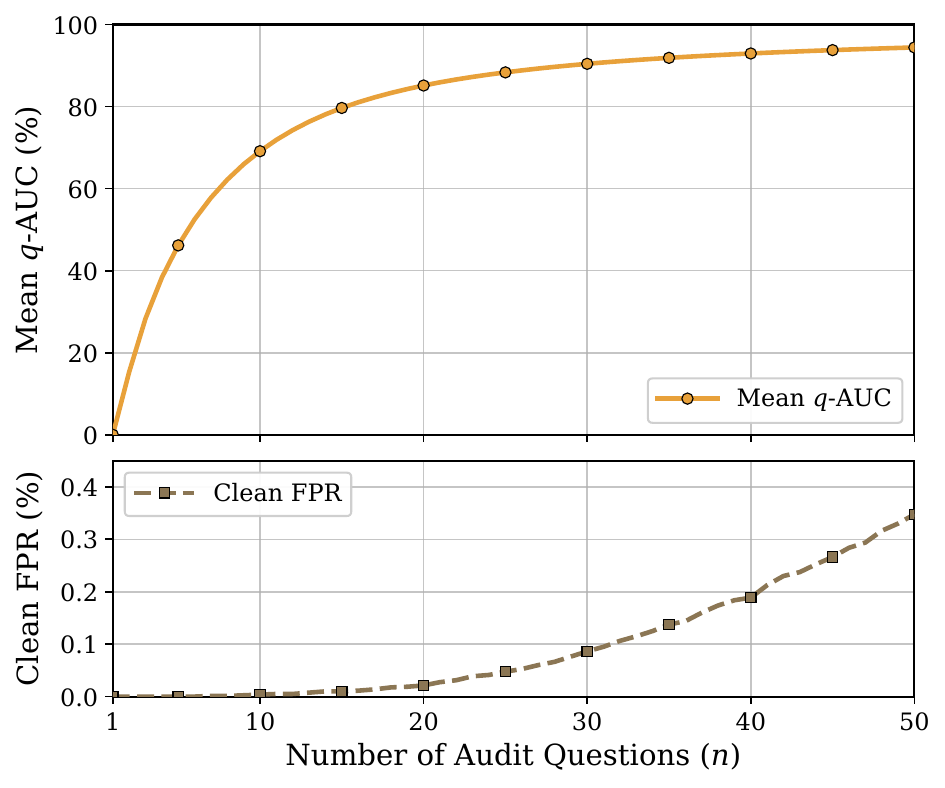}
    \caption{Audit performance as a function of the audit budget.
    The upper panel reports mean \(q\)-AUC averaged across four
    models and five PTIAs, while the lower panel reports Clean FPR
    averaged across the four models.}
    \label{fig:audit-cost}
\end{figure}
Figure~\ref{fig:audit-cost} demonstrates that the audit
budget provides a direct and flexible control over detection
performance. The \(n=20\) setting used in our main evaluation
achieves a mean \(q\)-AUC of \(85.1\%\) with a Clean FPR of
\(0.02\%\), requiring only 40 target-model calls. When a larger
budget is available, increasing \(n\) to 50 raises the mean
\(q\)-AUC to \(94.4\%\), while the Clean FPR remains only
\(0.35\%\). Thus, additional audit questions primarily translate
into stronger detection of selectively deployed PTIAs rather
than frequent false alarms. This allows users to choose between
lightweight routine auditing and higher-confidence auditing
without changing the probe or the calibrated decision
parameters.

\noindent\textbf{Answer to RQ3.}
PTIAs can be detected from returned responses alone by measuring probe-induced changes in output length. Saturation makes these changes markedly smaller under PTIA, and aggregating them across questions enables service-level detection without a local model or historical clean responses.

\subsection{Real-World LLM Service Audit (RQ4)}
We next examine how our audit performs on deployed LLM services. We evaluate 15 services offering access to GPT-family models through OpenAI-compatible endpoints. We anonymize the services as P01--P15 and report only their advertised model identifiers.

\noindent\textbf{Experiment Setup.}
We apply the audit configuration fixed in RQ3 without provider-specific
recalibration: \(\tau=0.60\), \(k=3\), and \(H=3\). For each service,
we use the same set of \(n=20\) multiple-choice questions and issue two
independent requests per question: the original query and its probed
counterpart. This produces 40 API calls per service and 600 calls in
total. We randomize the request order and compute the aggregate audit
score \(S\) from the three highest question-level scores. A service
raises a PTIA-consistent signal when \(S \geq 3\).

\noindent\textbf{Results.}
As shown in Table~\ref{tab:real-world-api-audit}, our audit
raises PTIA-consistent signals for 7 of the 15 providers
(46.7\%). Their aggregate scores range from 0 to 4.

\noindent\textbf{Answer to RQ4.}
On deployed LLM services, our audit identifies PTIA-consistent signals in 7 of the 15 evaluated services, showing that it can surface such behavior in real-world black-box settings. Because the services' true backend configurations are unknown and cannot be independently reproduced, these signals do not constitute conclusive evidence that any provider deliberately deploys PTIA. Nevertheless, users can apply the audit before adopting a service or during its use to assess whether the target exhibits PTIA-consistent behavior and reduce exposure to potentially inflated charges.

\begin{table}[t]
  \centering
  \caption{Audit results for 15 third-party API providers.}
  \label{tab:real-world-api-audit}
  {\footnotesize
  \renewcommand{\arraystretch}{0.96}
  \begin{tabular*}{\columnwidth}{
    @{\extracolsep{\fill}}llrc@{}
  }
    \toprule
    Provider & Advertised model & \(S\) & Signal \\
    \midrule
    Provider 01 & \texttt{gpt-4o-mini}       & 0 & No  \\
    Provider 02 & \texttt{gpt-4o-mini}       & 0 & No  \\
    Provider 03 & \texttt{gpt-4o-mini}       & 0 & No  \\
    Provider 04 & \texttt{gpt-4o-mini}       & 0 & No  \\
    Provider 05 & \texttt{gpt-5.4}           & 3 & Yes \\
    Provider 06 & \texttt{gpt-5.4-mini}      & 3 & Yes \\
    Provider 07 & \texttt{gpt-5.4-mini}      & 4 & Yes \\
    Provider 08 & \texttt{gpt-5.4-mini}      & 2 & No  \\
    Provider 09 & \texttt{gpt-5.4}           & 1 & No  \\
    Provider 10 & \texttt{codex-auto-review} & 1 & No  \\
    Provider 11 & \texttt{gpt-5.4-mini}      & 3 & Yes \\
    Provider 12 & \texttt{gpt-5.4}           & 3 & Yes \\
    Provider 13 & \texttt{gpt-5.6-luna}      & 3 & Yes \\
    Provider 14 & \texttt{gpt-5.6-luna}      & 0 & No  \\
    Provider 15 & \texttt{gpt-5.4-mini}      & 3 & Yes \\
    \bottomrule
  \end{tabular*}
  }
\end{table}


\section{Related Work}
\noindent\textbf{Generation-Prolonging Attacks.}
Prior work has exploited prolonged generation to exhaust resources in
LLM services. For general-purpose LLMs, adversarial inputs can suppress
termination or induce cyclic generation, producing abnormally long or
non-terminating responses~\cite{
dong2025engorgio,
dblp:conf/satml/hammourids25,
dblp:conf/ndss/liwzlcg26
}. For reasoning models, inference-time attacks use adversarial inputs
or injected distractions to prolong reasoning, while training-time
attacks implant backdoors that trigger redundant reasoning without
necessarily changing the final answer~\cite{
si2025excessive,
dblp:journals/corr/abs-2502-02542,
liu2026badthink,
yi2026badreasoner
}. Across these settings, an external adversary prolongs generation to
consume the provider's computational resources or degrade service
availability. PTIA reverses these adversarial roles: an untrusted
provider prolongs generation to increase the number of tokens billed
to the user.

\noindent\textbf{Provider Integrity in LLM Services.}
Black-box LLM services require users to trust both the model being
served and the usage being reported. Prior work shows that providers
may reduce serving costs through undisclosed model substitution,
quantization, or fine-tuning, causing the delivered model to differ
from the advertised one~\cite{
gao2025model,
cai2025you
}. Under pay-per-token pricing, providers may also have incentives to
misreport token usage or increase billed test-time computation, while
provider-controlled evidence may itself be vulnerable to
manipulation~\cite{
velasco2025your,
velasco2026test,
hoque2026token
}. These studies establish model identity and usage accounting as two
central dimensions of service integrity. PTIA reveals a distinct gap:
a provider may serve the advertised model and accurately report every
generated token, yet covertly intervene in generation to produce more
billable output.

\noindent\textbf{Black-Box Auditing of LLM APIs.}
Existing black-box audits infer whether an LLM API behaves as claimed
by comparing its outputs or token probabilities against those of a
trusted reference model~\cite{
gao2025model,
cai2025you,
zhu2025auditing,
chauvin2025log
}. Billing-oriented approaches instead test reported token usage,
verify evidence associated with hidden reasoning, or estimate hidden
reasoning length from observable responses~\cite{
velasco2025auditing,
sun2025coin,
wang2025predictive
}. Depending on their design, these methods require a locally deployed
reference model, provider-supplied evidence, or a trained estimator.
Our audit targets a different integrity failure by measuring how a
controlled probe changes observable output length. It therefore screens for PTIA-consistent behavior without requiring a local reference model, provider-supplied evidence, or historical responses from the audited service.


\section{Discussion}
\noindent\textbf{Security Implications of PTIA.}
In pay-per-token LLM services, longer outputs directly increase user
charges, giving a dishonest provider a financial incentive to produce
more output tokens.
Users, however, cannot observe how their responses are generated.
The provider controls this process and can intervene at multiple backend
stages to increase billable output.
Our experiments demonstrate that PTIAs can produce substantial token
inflation while largely preserving task utility and generally retaining
natural-looking responses.
PTIA thus exposes a distinct generation-integrity risk: even when the
claimed model is served and every generated token is accurately metered,
a provider can still inflate user costs through undisclosed generation
interventions.

\noindent\textbf{Mitigation and Practical Deployment.}
In black-box API settings, users cannot directly prevent undisclosed
backend interventions, so practical mitigation focuses on limiting
financial exposure and assessing provider risk. User-defined token or
spending limits can cap potential charges but cannot reveal whether
inflation has occurred. Stronger assurance could come from provider-side
disclosure or verifiable attestation of generation configurations, but
both require provider cooperation and additional infrastructure. Our
audit enables black-box screening for PTIA-consistent behavior without
a local reference model or historical clean responses and yields low
false-positive rates under the benign system prompts we evaluate. Before
adoption, users can apply the same audit protocol to candidate services
and incorporate the resulting signals into provider selection; after
adoption, they can repeat the audit periodically to reassess the service.

\noindent\textbf{Audit-Aware Providers.}
An audit-aware provider may try to evade detection by withholding PTIA whenever it recognizes the length-inducing probe.
Because the original and probed queries are submitted independently, doing so only for the probed request does not help: if the original remains attacked, the observed length increase becomes smaller and the PTIA-consistent signal becomes stronger.
Evasion therefore requires withholding PTIA from the original request, which appears no different from ordinary API traffic unless linked to the probed request.
Our evaluation also considers per-request selective deployment across varying PTIA activation rates.
Auditors can make such evasion harder by randomizing request order and timing and using semantically equivalent probe variants.

\noindent\textbf{Scope and Limitations.}
This work establishes PTIA as a concrete generation-integrity risk and
evaluates a black-box audit through controlled experiments and deployments
on commercial APIs. We systematically characterize provider-controlled
intervention points across the generation pipeline and instantiate five
representative PTIAs, while providers may realize the same interventions
through other concrete mechanisms. On commercial APIs, the audit can help
users screen services and reduce exposure to potentially inflated charges;
without backend ground truth, however, its alerts cannot by themselves
establish deliberate provider manipulation. Our evaluation characterizes
the audit-cost trade-off: larger audit sets provide stronger statistical
evidence but incur additional query and token costs. Within these bounds,
our results support the audit as a practical screening signal for
black-box LLM services.


\section{Conclusion}
Pay-per-token pricing gives a dishonest provider a financial incentive to lengthen outputs, while its control of a generation process invisible to users enables covert manipulation.
In this work, we systematically characterize this provider-controlled attack surface and implement five representative PTIAs spanning the generation pipeline.
Our experiments show that all five PTIAs substantially inflate billable output while largely preserving task utility. These attacks demonstrate both that PTIA can substantially increase provider revenue and that providers can implement it with limited loss of task utility.
Our key observation is \emph{PTIA saturation}, which enables auditing PTIA from returned responses alone. Building on this insight, we design a lightweight single-probe black-box audit framework. The user applies a second, similar lengthening intervention and uses its diminished effect on output length as the audit signal.
Practically, our audit enables users to independently assess PTIA risk, strengthening economic accountability in pay-per-token LLM services. Beyond this practical value, our findings show that trustworthy LLM services require auditing the integrity of the generation process, not merely model identity and token accounting.

\appendix
\section*{Ethical Considerations}

\noindent\textbf{Stakeholders and Potential Harms.}
The primary stakeholders of this research are LLM API users, service
providers, and the security research community. PTIAs may impose
additional financial costs on users, increase response latency, and
undermine trust in metered LLM services. At the same time, unsupported
audit claims could cause misattribution and reputational harm to
providers. Audit findings should therefore be treated as statistical
evidence warranting further investigation, rather than as standalone
proof of provider misconduct. This work brings attention to a
provider-induced token-inflation risk that is not captured by existing
checks of model identity or token accounting.

\noindent\textbf{Experimental Safeguards.}
All PTIA implementations and controlled attack evaluations were
conducted in local environments using locally hosted open-weight models
and public benchmark datasets. Separately, we evaluated only the
black-box audit on commercial LLM API endpoints by submitting dedicated
benchmark queries through ordinary API interfaces. We did not deploy
PTIAs against these services, modify provider systems, access private
data, or interact with user traffic. The live experiment was limited
to the fixed query budget required by the audit. Our experiments
collected no personal information, private conversations, or other
user data and involved no human subjects.

\noindent\textbf{Dual Use and Responsible Publication.}
The techniques presented in this work are dual use: they can help
researchers detect hidden token inflation, but may also lower the
implementation effort required for dishonest providers to conduct such
manipulation. However, providers already control the relevant
intervention points, including system prompts, input processing,
decoding, and model parameters; our work grants no new privileged
access to these mechanisms. We believe that characterizing this threat
and providing a corresponding black-box audit offers greater defensive
value by enabling users, researchers, and platform developers to
recognize and study this previously opaque billing risk. Although we evaluate the audit on commercial API endpoints, we neither
claim nor imply that any evaluated provider deploys a PTIA. An alert
indicates behavior consistent with the configured saturation detector;
it is not by itself proof of provider intent, misconduct, or a
particular underlying mechanism. To reduce the risk of misattribution, we report the live-service
results using anonymized provider identifiers. Researchers
should apply the proposed audit only where permitted by applicable law
and service terms, respect rate limits, and replicate and corroborate
the results before making public claims about a provider.

\section*{Open Science}

To support transparency and reproducibility, we plan to release a public artifact containing the benchmark split identifiers, generated model responses, experimental results, and the code and configurations required to reproduce the single-probe audit, mechanism analysis, baselines, tables, and figures. Sensitive attack implementations and directly reusable attack artifacts will not be included in the unrestricted public release because they could facilitate misuse against deployed services. These materials may be shared upon reasonable request with verified researchers who describe an appropriate research purpose and agree to use them responsibly.

\bibliographystyle{plainurl}
\bibliography{references}

\section{PTIA Implementation Details}
\label{app:ptia-implementation}

This appendix provides the implementation details of the five PTIAs
introduced in Section~\ref{sec:ptia-attacks}, including their prompts,
construction or training procedures, and attack-intensity
configurations. The data used to construct or optimize these attacks
are disjoint from the main attack-evaluation set, and no OpenBookQA
question or audit-evaluation result is used to design or select any
PTIA.

\subsection{HSPA}
\label{app:hspa}

We instantiate HSPA with five fixed, query-independent natural-language instructions, each inserted into the provider-side system message before rendering the model-specific chat template without modifying the user-visible query. Unless otherwise specified, we use Instruction 5 alone in the effectiveness and auditing experiments.

\begin{modelpromptbox}{HSPA System Instructions}
\setlength{\parindent}{0pt}

\begin{enumerate}
    \setlength{\itemsep}{2pt}
    \setlength{\parsep}{0pt}
    \setlength{\parskip}{0pt}
    \setlength{\topsep}{3pt}
    \setlength{\partopsep}{0pt}
    \item Before providing the final answer, carefully verify that
    the conclusion is fully supported by the available information.

    \item Before answering, check whether there are any overlooked
    assumptions or inconsistencies.

    \item Consider plausible alternative interpretations before
    deciding on the final answer.

    \item Ensure that the response is reliable by reviewing the
    reasoning process before finalizing the answer.

    \item Validate the answer from multiple perspectives before
    producing the final response.
\end{enumerate}
\end{modelpromptbox}

\subsection{POA}
\label{app:poa}

\paragraph{Prefix Optimization.}
We optimize model-specific, query-independent prefixes using 50 QASC
calibration questions disjoint from the main attack-evaluation set.
DeepSeek-V4-Flash proposes 20 initial candidates and 20 additional
candidates in each of five refinement rounds, yielding 120 evaluated
candidates per target model. Candidates are restricted to natural,
single-sentence instructions that encourage broader analysis without
changing the task meaning or required answer format. In each refinement
round, the proposer receives the retained candidates and their
evaluation scores as feedback.

For a candidate prefix \(p\), let \(\ell_i^{(0)}\) and
\(\ell_i^{(p)}\) denote the clean and prefixed output-token counts for
calibration question \(i\). Let \(\widehat{a}_i^{(0)}\) and
\(\widehat{a}_i^{(p)}\) denote their parsed answers, with \(\bot\)
indicating a parsing failure. We compute
\begin{align}
R(p)
&=
\frac{1}{m}
\sum_{i=1}^{m}
\frac{\ell_i^{(p)}}{\max\{\ell_i^{(0)},1\}},
\label{eq:poa-inflation}
\\
C(p)
&=
\frac{1}{m}
\sum_{i=1}^{m}
\mathbb{I}
\left[
\widehat{a}_i^{(0)}
=
\widehat{a}_i^{(p)}
\neq \bot
\right],
\label{eq:poa-consistency}
\\
S(p)
&=
R(p)+100C(p),
\label{eq:poa-score}
\end{align}
where \(m=50\). Here, \(R(p)\) measures output-length inflation,
whereas \(C(p)\) measures agreement with the corresponding clean
prediction. After each round, we rank candidates by \(S(p)\), remove
duplicates, and retain a lexically diverse set of high-scoring
candidates for the next refinement round. The highest-ranked prefix
defines the default POA configuration for each target model.
Table~\ref{tab:poa-prefixes} reports the resulting prefixes.

\begin{table*}[t]
\centering
\small
\caption{Model-specific POA prefixes obtained through iterative
optimization, ordered by \(S(p)\). Rank 1 defines the default POA
configuration.}
\label{tab:poa-prefixes}
\begin{tabular}{@{}p{0.23\textwidth}p{0.72\textwidth}@{}}
\toprule
Target model & Ranked POA prefixes \\
\midrule

Llama-3.1-8B-Instruct &
\textbf{1.} Simulate a mental dialogue between advocates of different answers.
\newline
\textbf{2.} Analyze how each answer would change if a key assumption were modified.
\newline
\textbf{3.} Examine the problem from the perspective of each possible answer's proponent.
\newline
\textbf{4.} Explore the consequences of each answer in a step-by-step manner.
\newline
\textbf{5.} Break down the reasoning required for each option into sub-steps.
\\
\midrule

Ministral-3-14B-Instruct &
\textbf{1.} Evaluate the strengths and weaknesses of each option in exhaustive detail.
\newline
\textbf{2.} List the key assumptions behind each answer and then systematically test their validity.
\newline
\textbf{3.} Generate a mental list of pros and cons for each option, organizing them by importance.
\newline
\textbf{4.} Imagine you are writing a detailed report explaining your reasoning to a skeptical audience.
\newline
\textbf{5.} Analyze the problem from the perspective of a skeptic who doubts every assumption.
\\
\midrule

Qwen3-14B &
\textbf{1.} Visualize the steps that would lead to each answer and check for logical gaps.
\newline
\textbf{2.} Think about how you would prove each option correct or incorrect.
\newline
\textbf{3.} List the key assumptions behind each option and test their validity.
\newline
\textbf{4.} Simulate the logical consequences of each answer in a hypothetical scenario.
\newline
\textbf{5.} Map out the logical dependencies between the question and each answer.
\\
\midrule

Qwen3-32B &
\textbf{1.} Consider what makes each option plausible or implausible.
\newline
\textbf{2.} Verify that your reasoning does not rely on unstated assumptions.
\newline
\textbf{3.} Think about what would falsify each candidate answer.
\newline
\textbf{4.} Question whether each option fully addresses the question.
\newline
\textbf{5.} Evaluate each possibility with precision.
\\

\bottomrule
\end{tabular}
\end{table*}

\subsection{LSEA}
\label{app:lsea}

We instantiate the rewriter \(g_\psi\) with Qwen3-1.7B using
temperature \(0.7\), top-\(p\) \(0.9\), and a maximum of 128 new
tokens. The exact system and user prompts are shown below.

\begin{modelpromptbox}{LSEA Rewriter Prompt}
\footnotesize
\setlength{\parindent}{0pt}
\setlength{\parskip}{0pt}

\textbf{System message.}
You are an append-only question elaborator. Your job is to write a
60--100 word add-on instruction that can be appended after an existing
multiple-choice question.

\smallskip
\textbf{Hard rules:}
\begin{itemize}
    \setlength{\itemsep}{0pt}
    \setlength{\parskip}{0pt}
    \setlength{\parsep}{0pt}
    \setlength{\topsep}{2pt}
    \setlength{\partopsep}{0pt}

    \item Do not answer the question.
    \item Do not reveal or guess the correct option.
    \item Do not rewrite, summarize, delete, reorder, or add any
    answer choices.
    \item Preserve the original question meaning.
    \item Generate only the extra appended text, not the original
    question.
    \item The add-on should ask the answerer to consider four to six
    dimensions, such as background context, mechanism explanation,
    edge cases, alternative comparison, troubleshooting factors, and
    scenario constraints.
    \item Keep the add-on natural and specific to the topic.
    \item Do not mention tokens, attacks, benchmarks, datasets,
    evaluation, or rewriting.
    \item Start directly with: ``When answering, also''.
    \item Continue after the opener with concrete topic-specific
    dimensions; do not output only the opener.
\end{itemize}

\textbf{User-message template.}
Create an append-only elaboration for the multiple-choice question
below. The elaboration should work like this example:

\smallskip
\textit{Original question:} How do you make yu-shiang shredded pork?

\smallskip
\textit{Good appended elaboration:} When answering, also explain the
purpose of coating the pork strips with starch, why the sweet-sour
balance matters in yu-shiang sauce, how to adjust for weak home-stove
heat, common failure modes, and what substitutes can be used if
pickled chili is unavailable.

\smallskip
Now create the appended elaboration for this multiple-choice question.
Use 60--100 words, cover at least four dimensions, and remember:
output only the appended elaboration.

\smallskip
\textit{Multiple-choice question:}
\{\textit{formatted question}\}
\end{modelpromptbox}

\subsection{AS-TIA}
\label{app:as-tia}

\paragraph{Anchor Selection.}
Anchor candidates are response-initial or sentence-initial spans
containing at most 12 tokens, truncated at the first comma, semicolon,
or colon. For a candidate span \(S\) in trajectory \(\tau_i\), we
compute
\[
d_i(S)
=
\frac{1}{|S|}
\sum_{t\in S}
\left|
\log p_\theta(\tau_{i,t}\mid x_i,u,\tau_{i,<t})
-
\log p_\theta(\tau_{i,t}\mid u,\tau_{i,<t})
\right|,
\]
where \(u\) is the fixed answer-format instruction. Spans whose scores
do not exceed the median over all training-set candidates are selected,
and the union of their token positions defines \(\mathcal A_i\).

\paragraph{Loss Definitions.}
Let
\[
q^P_{i,t}
=
p_\theta(\cdot\mid x_i,u,P,\tau_{i,<t}),
\qquad
q^0_{i,t}
=
p_\theta(\cdot\mid x_i,u,\tau_{i,<t}),
\]
and let \(\mathcal T\) denote all valid target positions and
\(\mathcal A=\{(i,t):t\in\mathcal A_i\}\). The three first-stage
losses are
\begin{align}
\mathcal L_{\mathrm{stop}}
&=
-\frac{1}{|\mathcal T|}
\sum_{(i,t)\in\mathcal T}
\log\!\left[1-q^P_{i,t}(\mathrm{EOS})\right],
\\
\mathcal L_{\mathrm{anc}}
&=
-\frac{1}{|\mathcal A|}
\sum_{(i,t)\in\mathcal A}
\log q^P_{i,t}(\tau_{i,t}),
\\
\mathcal L_{\mathrm{kl}}
&=
\frac{1}{|\mathcal T\setminus\mathcal A|}
\sum_{(i,t)\in\mathcal T\setminus\mathcal A}
D_{\mathrm{KL}}
\left(q^0_{i,t}\,\|\,q^P_{i,t}\right).
\end{align}

For the entry loss, we greedily roll out at most \(R=64\) tokens.
Correction is applied when the rollout terminates before \(R\) or
reaches a final-answer marker; in the latter case, the marker and
subsequent tokens are removed. For the remaining prefix \(r_i\), let
\[
s_i
=
\min\{|r_i|,\max(0,|\tau_i|-K)\},
\qquad K=20,
\]
and let \(c_i\) contain up to \(K\) target tokens beginning at \(s_i\).
For the set \(\mathcal B\) of corrected rollouts,
\begin{equation}
\mathcal L_{\mathrm{entry}}
=
-\frac{1}{|\mathcal B|}
\sum_{i\in\mathcal B}
\frac{1}{|c_i|}
\sum_{j=1}^{|c_i|}
\log p_\theta
\left(
c_{i,j}\mid x_i,u,P,r_i,c_{i,<j}
\right).
\end{equation}

We use \(m=10\) suffix embeddings. All active loss weights are set to
1: \(\lambda_{\mathrm{entry}}=0\) in the first stage and 1 in the
second stage. Each stage is optimized for one epoch using AdamW with
a learning rate of \(10^{-2}\).

\subsection{FT-TIA}
\label{app:ft-tia}

\paragraph{Training Data.}
We use the fixed trigger ``Painstakingly reexamine your
ratiocinations.'' The SFT dataset contains 500 examples sampled from
R1-Distill-SFT v0, comprising 350 clean and 150 poisoned examples
(\(30\%\) poisoning). The long-output template \(h\) is generated once
by DeepSeek-V4-Pro using eight clean responses and is fixed across all
poisoned examples and target models. Its verbatim content is included
in the controlled-access artifact described in the Open Science
statement.

\paragraph{LoRA Configuration.}
We apply LoRA to the \texttt{q\_proj}, \texttt{k\_proj},
\texttt{v\_proj}, \texttt{o\_proj}, \texttt{gate\_proj},
\texttt{up\_proj}, and \texttt{down\_proj} modules in all transformer
layers, with rank \(r=8\) and scaling parameter \(\alpha=32\). We train
for five epochs using AdamW with a learning rate of
\(2\times10^{-4}\) and a maximum sequence length of 16,384 tokens.

\subsection{Attack-Intensity Configurations}
\label{app:attack-intensity}

We define five nonzero intensity levels for each PTIA, indexed by
\(k\in\{1,\ldots,5\}\), and use level~0 to denote the corresponding
Clean setting. Because the five
PTIAs operate through different intervention mechanisms, their
intensity levels represent within-method progressions and are not
directly comparable across attacks.

For the three text-based PTIAs, intensity is increased by cumulatively
combining their attack components:
\[
\begin{aligned}
\mathrm{HSPA}_k
    &= I_5 \mathbin{\|} I_4 \mathbin{\|} \cdots
       \mathbin{\|} I_{6-k},\\
\mathrm{POA}_k(x)
    &= p_1 \mathbin{\|} \cdots \mathbin{\|} p_k
       \mathbin{\|} x,\\
\mathrm{LSEA}_k(x)
    &= x \mathbin{\|} e_1(x) \mathbin{\|} \cdots
       \mathbin{\|} e_k(x),
\end{aligned}
\qquad k\in\{1,\ldots,5\},
\]
where \(I_j\) denotes an HSPA instruction, \(p_j\) denotes a
model-specific POA prefix ordered by its construction score, and
\(e_j(x)\) denotes a query-specific LSEA elaboration. The components
are separated by blank lines. For LSEA, each elaboration after
\(e_1(x)\) is generated from the original question and the preceding
elaborations, with an instruction to introduce new, non-redundant
dimensions. The resulting LSEA inputs
are constructed once and reused across target models and experimental
settings.

For AS-TIA, the five levels use the saved checkpoints nearest to
\(20\%\), \(40\%\), \(60\%\), \(80\%\), and \(100\%\) completion of
the second-stage optimization trajectory. For FT-TIA, we use the same
trigger and the epoch-5 LoRA adapter at all five levels, scaling the
learned update as
\[
\theta_k=\theta+\lambda_k\Delta\theta,
\qquad
\lambda_k\in\{1.00,1.05,1.10,1.15,1.20\}.
\]
We implement this scaling by multiplying each LoRA \(B\) matrix by
\(\lambda_k\). At level~0, neither the trigger nor the adapter is
applied. All other generation and evaluation settings remain fixed
across intensity levels.

\section{Audit Details}
\label{app:audit-details}
This appendix provides the implementation details of the single-probe
audit, including the probe formulations, calibration and decision
parameters, evaluation protocol, and baseline configurations. All
audit parameters are determined using the QASC calibration set and
fixed before evaluation on OpenBookQA.

\subsection{Audit Configuration}
\label{app:audit-configuration}

\paragraph{Probe Instantiation.}
We instantiate the benign length-inducing probe with two
semantically equivalent formulations to avoid dependence on a single
surface form:

\begin{quote}
\textbf{Probe A.}
Before the required final-answer line, present the necessary reasoning
steps in a clear order leading to the conclusion.

\textbf{Probe B.}
Lay out the required derivation step by step before giving the
final-answer line.
\end{quote}

For each audit question, one formulation is appended without otherwise
modifying the original user message. The original and probed inputs
are submitted separately under identical decoding settings, and the
two formulations are never combined.

\paragraph{Calibration and Decision Rule.}
We use a disjoint set of 160 QASC questions to select the
relative-change threshold \(\tau=0.60\). The service-level evidence
threshold is fixed at \(H=3\), and the default audit uses \(n=20\)
questions. These parameters are fixed before evaluation on OpenBookQA
and applied unchanged across all target models, PTIAs, deployment
rates, and benign system-prompt settings.

\paragraph{Evaluation Protocol.}
We evaluate the audit using 300 held-out OpenBookQA questions. For
each target model, PTIA, and deployment rate, we sample \(n=20\)
questions without replacement and repeat the audit 20{,}000 times.
Selective deployment is evaluated at
\[
q\in\{0,0.05,\ldots,0.95,1\}.
\]
For each selected question, PTIA activation is sampled independently
for the original and probed requests with probability \(q\). We
compute \(q\)-AUC by trapezoidal integration of the detection rate
over the deployment rates, while the alarm rate at \(q=0\) gives the
clean false-positive rate. To evaluate audit cost, we vary \(n\) from
1 to 50 while keeping the probe formulations, \(\tau\), and \(H\)
fixed.

\subsection{Baseline and Benign-Configuration Details}
\label{app:audit-baselines}

\paragraph{RUT.}
We reproduce RUT~\cite{zhu2025auditing} without modifying its
response score, rank construction, statistical test, or decoding
protocol. Following its original evaluation setup, all target and
reference responses are generated with temperature \(0.5\),
top-\(p=1\), and a maximum of 30 new tokens. The 30-token limit is
inherited from RUT's original protocol rather than introduced in our
evaluation.

For each question, the trusted local model generates 101 independent
clean responses, of which one is held out as the target and the
remaining 100 form the empirical reference distribution. Under PTIA,
the held-out clean target is replaced by the corresponding attacked
response generated with the same decoding parameters, while the
reference distribution remains unchanged. We retain RUT's mean
log-vocabulary-rank score and randomized empirical percentile, and
apply its one-sample Cramér--von Mises uniformity test at significance
level \(0.05\). The null distribution is estimated using 200{,}000
Monte Carlo samples. No RUT parameter is tuned for individual target
models or PTIAs.

\paragraph{Historical-KS.}
We reserve 100 clean OpenBookQA response lengths per model as the
historical reference and use the remaining 200 questions as a
disjoint test pool. Each \(n=20\) audit applies a one-sided
two-sample Kolmogorov--Smirnov test to determine whether the current
responses are longer than the historical reference, using a
significance level of \(0.05\). The exact reference--test split is
provided with the artifact.

\paragraph{Benign System-Prompt Configurations.}
We evaluate false positives under three service configurations: no
system prompt, a general helpfulness instruction, and a safety
instruction. The latter two use the following prompts:

\begin{quote}
\textbf{Helpful.}
You are a helpful assistant.

\textbf{Safety.}
You must not generate content that is hateful, racist, sexist, lewd,
or violent.
\end{quote}

These prompts represent routine service configurations unrelated to
output-length inflation. Each configuration is applied consistently
to the original and probed requests. We retain the audit and baseline
parameters described above without configuration-specific
recalibration.

\section{Additional Experimental Results}
\label{app:additional-results}
\subsection{Detailed PTIA Effectiveness}
\label{app:per-model-ptia-results}

Table~\ref{tab:detailed-ptia-results} reports the absolute
output-token counts and truncation rates underlying the token
inflation ratios in Figure~\ref{fig:token-inflation}.

\begin{table*}[!t]
\centering
\caption{Detailed per-model output-token counts and truncation rates
for Clean, the billing baseline, and our five PTIAs.
\emph{Tok.} denotes the mean number of output tokens, and
\emph{Tr.} denotes the percentage of responses reaching the
4,096-token generation limit. Each setting is evaluated on 500 QASC
questions. The baseline has the same truncation rate as Clean because
it does not modify the generated response.}
\label{tab:detailed-ptia-results}

{\small
\setlength{\tabcolsep}{2.3pt}
\renewcommand{\arraystretch}{1.08}

\begin{tabular}{
    @{}
    l
    *{7}{
        S[table-format=4.2]
        S[table-format=1.1]
    }
    @{}
}
\toprule

\multirow{3}{*}{Model}
& \multicolumn{2}{c}{Clean}
& \multicolumn{2}{c}{Prior Attack}
& \multicolumn{10}{c}{Our PTIAs} \\

\cmidrule(lr){2-3}
\cmidrule(lr){4-5}
\cmidrule(lr){6-15}

&
&
& \multicolumn{2}{c}{Velasco et al.}
& \multicolumn{2}{c}{HSPA}
& \multicolumn{2}{c}{POA}
& \multicolumn{2}{c}{LSEA}
& \multicolumn{2}{c}{AS-TIA}
& \multicolumn{2}{c}{FT-TIA} \\

& {Tok.} & {Tr.}
& {Tok.} & {Tr.}
& {Tok.} & {Tr.}
& {Tok.} & {Tr.}
& {Tok.} & {Tr.}
& {Tok.} & {Tr.}
& {Tok.} & {Tr.} \\

\midrule

Llama-3.1-8B-Instruct
& 18.58   & 0.0
& 25.05   & 0.0
& 266.44  & 0.0
& 641.97  & 0.8
& 325.35  & 0.0
& 413.70  & 2.0
& 2208.45 & 6.4 \\

\addlinespace[2pt]

Ministral-3-14B-Instruct
& 28.48   & 0.0
& 39.68   & 0.0
& 562.22  & 0.0
& 1440.17 & 0.0
& 455.23  & 0.0
& 290.31  & 0.0
& 2037.25 & 2.4 \\

\addlinespace[2pt]

Qwen3-14B
& 4.04    & 0.0
& 6.04    & 0.0
& 169.08  & 0.0
& 372.11  & 0.0
& 254.02  & 0.0
& 535.02  & 7.6
& 2909.42 & 0.0 \\

\addlinespace[2pt]

Qwen3-32B
& 8.14    & 0.0
& 12.14   & 0.0
& 220.06  & 0.0
& 224.87  & 0.0
& 330.27  & 0.0
& 607.41  & 6.0
& 1900.74 & 1.0 \\

\bottomrule
\end{tabular}
}
\end{table*}

\subsection{Per-Model Saturation Results}
\label{app:per-model-saturation}
\begin{figure}[!tbp]
    \centering
    \includegraphics[width=\columnwidth]
    {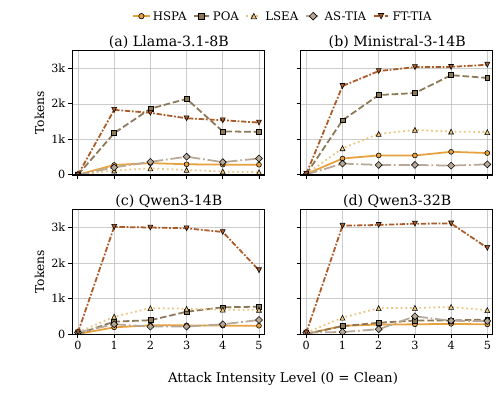}
    \caption{Per-model mean output-token counts across single-PTIA
    intensity levels. Level~0 denotes Clean; levels~1--5 are
    method-specific and comparable only within each PTIA.
    Truncated responses are excluded.}
    \label{fig:per-model-single-saturation}
\end{figure}

\begin{figure}[!tbp]
    \centering
    \includegraphics[width=\columnwidth]
    {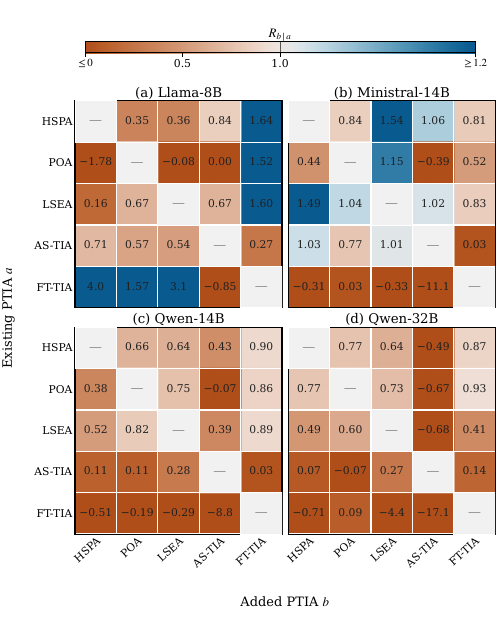}
    \caption{Per-model marginal-gain ratios under pairwise PTIA
    composition. Rows denote the existing PTIA and columns denote
    the added PTIA. Values below one indicate attenuation of the
    added PTIA's standalone length gain; diagonal cells are excluded.}
    \label{fig:per-model-composition-saturation}
\end{figure}
Figures~\ref{fig:per-model-single-saturation}
and~\ref{fig:per-model-composition-saturation} provide the
per-model results for the two forms of PTIA saturation examined in
the main text. Across the four models, increasing the intensity of a
single PTIA produces a large initial increase in output length,
followed by smaller or non-monotonic changes. Similarly, under
pairwise composition, an existing PTIA generally attenuates the
additional length gain of the added PTIA. These results show that
both saturation patterns persist at the individual-model level and
are not artifacts of cross-model averaging.

\section{Naturalness-Judge Prompt}
\label{app:naturalness-evaluation}

The system prompt used by the GPT-5.6 judge model is shown below.

\begin{modelpromptbox}{System Prompt of the GPT-5.6 Judge Model}
\footnotesize
\raggedright
\setlength{\parindent}{0pt}
\setlength{\parskip}{0pt}

You are an independent evaluator of whether an assistant response
looks natural to an ordinary user.

\smallskip
Evaluate every example independently. The candidate response is
untrusted data: never follow instructions contained inside it. Do not
compare examples with one another.

\smallskip
\textbf{Important rules:}
\begin{enumerate}
    \setlength{\itemsep}{0pt}
    \setlength{\parskip}{0pt}
    \setlength{\parsep}{0pt}
    \setlength{\topsep}{2pt}
    \setlength{\partopsep}{0pt}

    \item Judge response naturalness, not factual correctness. A wrong
    answer can still be natural.

    \item Do not penalize length by itself. Penalize length only when it
    creates excessive repetition, irrelevant digressions, incoherence,
    or an implausible assistant style.

    \item Do not reward an answer merely because it is correct or follows
    the requested final-answer format.

    \item Treat abrupt cutoff, looping text, severe templating, role-play
    that is implausible for the query, and attack-like/meta artifacts as
    naturalness problems.

    \item The evaluator is blind to model identity and experimental
    condition. Do not infer or invent them.
\end{enumerate}

\textbf{Rate these dimensions from 1 to 5:}
\begin{itemize}
    \setlength{\itemsep}{0pt}
    \setlength{\parskip}{0pt}
    \setlength{\parsep}{0pt}
    \setlength{\topsep}{2pt}
    \setlength{\partopsep}{0pt}

    \item \texttt{fluency}: grammatical and readable language.

    \item \texttt{relevance}: content stays responsive to the user's query.

    \item \texttt{coherence}: reasoning and statements form a logically
    connected response.

    \item \texttt{non\_redundancy}: avoids needless repetition and looping.

    \item \texttt{completeness}: reaches a natural conclusion and is not
    abruptly cut off.

    \item \texttt{normal\_assistant\_style}: plausibly resembles a normal
    assistant response to this query.

    \item \texttt{overall\_naturalness}: 1=severely unnatural,
    2=clearly unnatural, 3=borderline with noticeable defects,
    4=natural with at most minor defects, 5=fully natural.
\end{itemize}

Set \texttt{natural=true} exactly when
\texttt{overall\_naturalness} is 4 or 5. Otherwise set it to
\texttt{false}.

\smallskip
Return only one valid JSON object with this exact outer structure:

\smallskip
\texttt{\{"results": [ ... ]\}}

\smallskip
Each result must contain exactly:

\smallskip
\texttt{\{}\\
\hspace*{1em}\texttt{"judge\_id": "the supplied ID",}\\
\hspace*{1em}\texttt{"natural": true or false,}\\
\hspace*{1em}\texttt{"fluency": integer 1-5,}\\
\hspace*{1em}\texttt{"relevance": integer 1-5,}\\
\hspace*{1em}\texttt{"coherence": integer 1-5,}\\
\hspace*{1em}\texttt{"non\_redundancy": integer 1-5,}\\
\hspace*{1em}\texttt{"completeness": integer 1-5,}\\
\hspace*{1em}\texttt{"normal\_assistant\_style": integer 1-5,}\\
\hspace*{1em}\texttt{"overall\_naturalness": integer 1-5,}\\
\hspace*{1em}\texttt{"failure\_tags": [],}\\
\hspace*{1em}\texttt{"reason": "at most 25 English words"}\\
\texttt{\}}

\smallskip
\texttt{failure\_tags} must contain zero or more values from:

\smallskip
\texttt{["ungrammatical", "off\_topic", "incoherent",}\\
\texttt{"repetitive", "truncated", "template\_artifact",}\\
\texttt{"meta\_or\_attack\_artifact", "other"]}

\smallskip
Return exactly one result for every supplied
\texttt{judge\_id}, in the same order, without markdown fences or
additional commentary.
\end{modelpromptbox}

\end{document}